\documentstyle[12pt]{article}
\input epsf.tex

\newcommand{\beq}{\begin{equation}}
\newcommand{\eeq}{\end{equation}}
\newcommand{\beqa}{\begin{eqnarray}}
\newcommand{\eeqa}{\end{eqnarray}}
\newcommand{\beqar}{\begin{eqnarray*}}
\newcommand{\eeqar}{\end{eqnarray*}}

\newcommand{\inn}{\!\cdot\!}

\renewcommand{\l}{\lambda}
\renewcommand{\L}{\Lambda}

\newcommand{\z}{\zeta}

\newcommand{\eg}{{\it e.g.,}\ }
\newcommand{\ie}{{\it i.e.,}\ }
\newcommand{\labell}[1]{\label{#1}} 
\newcommand{\reef}[1]{(\ref{#1})}
\newcommand\prt{\partial}

\newcommand\Tr{{\rm Tr}}
\newcommand\M[2]{M^{#1}{}_{#2}}
\begin{document}

\vspace*{1cm}

\begin{center}
{\bf \Large Non-abelian expansion of  S-matrix elements and
\\non-abelian  tachyon DBI action
  \\

 }
\vspace*{1cm}

{Kazem Bitaghsir-Fadafan and
Mohammad R. Garousi}\\
\vspace*{0.2cm}
{ Department of Physics, Ferdowsi university, P.O. Box 1436, Mashhad, Iran}\\
\vspace*{0.1cm}
and\\
{ Institute for Studies in Theoretical Physics and Mathematics
IPM} \\
{P.O. Box 19395-5531, Tehran, Iran}\\
\vspace*{0.4cm}

\vspace{2cm}
ABSTRACT
\end{center}
We apply the prescription proposed in hep-th/0307197 for
non-abelian expansion of  S-matrix elements, to the S-matrix
element of four tachyons and one gauge field  in superstring
theory. We  show that the leading order terms of the expansion are
in perfect agreement with the  non-abelian generalization of the
tachyon DBI action in which  the tachyon potential is
$V(T)=1+\pi\alpha' m^2T^2+\frac{1}{2!}(\pi\alpha'
m^2T^2)^2+\cdots$ where $m^2=-1/(2\alpha')$ is the mass of
tachyon. This calculation fixes the coefficient of four-tachyon
couplings  without on-shell ambiguity.

\vfill
\setcounter{page}{0}
\setcounter{footnote}{0}
\newpage

\section{Introduction} \label{intro}
Decay of unstable branes is an interesting process which might
shed new light in understanding properties of string theory in
time-dependent backgrounds \cite{mgas}-\cite{NJ}. In particular,
by studying the unstable branes in the boundary conformal field
theory (BCFT), Sen has shown that the end of tachyon rolling  in
this theory is a tachyon matter with zero pressure and non-zero
energy density \cite{asen2}. These results can be  reproduced in
field theory using the tachyon DBI action \cite{mg,ebmr}. See
\cite{ASen} for review.

The tachyon DBI action was originally proposed as an action which
is consistent with the non-commutative expansion of the S-matrix
elements involving open string tachyons \cite{mg}. The example
considered in \cite{mg} was the S-matrix element of one graviton
and two tachyons in the presence of background B-flux. Using the
fact that the field theory on the world-volume of D-brane with
B-flux is a non-commutative theory \cite{AC}-\cite{NS}, one
naturally  expects there should be an expansion for the S-matrix
element whose first leading order term which is a massless pole is
reproduced by the non-commutative kinetic term of the tachyon
action. Such expansion which is not the usual $\alpha'$-expansion
has been found for the above S-matrix element in \cite{mg}. It is
an expansion around $s\rightarrow 0$ where the Mandelstam variable
is $s=-1/2-\alpha'k_1\cdot k_2$ where $k_1,k_2$ are the momenta of
the tachyons. It has been shown while the first leading order term
of the expansion is exactly reproduced by the non-commutative
kinetic term, the next order terms which are contact terms,
indicates that the kinetic term of tachyon can not be in the
standard form. It should appear, in fact,  in the tachyon DBI form
\cite{mg}\footnote{Throughout the paper, we are appealing to the
specific meaning of  the tachyon action as a generating functional
for producing the leading terms of the string theory S-matrix
elements.}.

When there are  $N$ coincident D-branes, the $U(1)$ gauge symmetry
of an individual D-brane is enhanced to the non-abelian $U(N)$
symmetry  \cite{ew1}. In this case, one expects  there should be
an expansion for any string theory S-matrix element whose first
leading order terms are reproduced by  the non-abelian gauge
invariant kinetic term of the tachyon action. This expansion has
been called in \cite{garousi} the non-abelian expansion of the
S-matrix elements. As an example for this case, the non-abelian
expansion of the S-matrix element of four tachyons has been found
in \cite{garousi}. This expansion is not the usual
$\alpha'$-expansion. Like the S-matrix element of four massless
transverse scalars, the contact terms of the expansion are ordered
in terms of $\z(2)$, $\z(3)$, $\cdots$. However, at each
$\z(n)$-order, there are terms at different $\alpha'$ order. It
has been shown that while the first leading order terms of the
expansion which are massless poles are reproduced by the
non-abelian kinetic term, the next leading order terms which are
contact terms are exactly reproduced by the non-abelian
generalization of tachyon DBI action in which the tachyon
potential is $V(T)=1+\pi\alpha' m^2T^2+\frac{1}{2!}(\pi\alpha'
m^2T^2)^2+\cdots$ where $m^2$ is mass of tachyon. Since these
couplings are fixed by comparing the contact terms of the S-matrix
element, there is the on-shell ambiguity in calculating these
terms, \eg the coefficient of $T^4$ in the tachyon potential has
the ambiguity that  one can not distinguish between, say, $T^4$
and $T^3\prt^2T$. Both have the same contribution to the S-matrix
element of four tachyons. This on-shell ambiguity  can be fixed by
studying  the S-matrix elements in which one of the tachyon
appears as off-shell in the Feynman amplitude, \eg the S-matrix
element of four tachyons and one gauge field.

The non-commutative/non-abelian expansion is not
$\alpha'$-expansion, so one may conclude that the non-leading
terms of the expansion are not related to higher derivative
correction to tachyon DBI action\footnote{See \cite{DKV}, for
discussion on validity of tachyon DBI action as effective theory
of non-BPS D-brane for spacial slowly  varying tachyon field in
tachyon rolling background in BSCT. There is a natural $\alpha'$
expansion for the tachyon S-matrix elements here, however,
expansion is in terms of spatial momentum of tachyons.}. However,
using the fact that for on-shell tachyon $\alpha'\prt_a\prt^a
T\sim T$, the higher derivatives of tachyon action at a fixed
$\alpha'$ order may produce contact terms at various $\alpha'$
order in the S-matrix element. Moreover, if a tachyon potential
multiplies a higher derivative term, then higher derivative terms
at a fixed number of derivative of tachyon \eg $\prt\prt T$-order,
produce different $\alpha'$ order in the S-matrix element, \eg
$\alpha'^2T^2\prt_a\prt_b T\prt^a\prt^b T$ and
$\alpha'^3\prt_cT\prt^cT \prt_a\prt_b T\prt^a\prt^b T$ are both at
second-order derivative of tachyon, however they produce different
$\alpha'$ order terms in the S-matrix element of four tachyons.
Hence, one can not argue  that the non-leading terms of the
expansion are not related to the higher derivative correction to
tachyon DBI action either. Using the fact that the S-matrix method
can not fix the coefficients of all gauge invariant structures
uniquely, \eg there is  a field redefinition freedom
\cite{AAT,AAT1}, one may correspond  the non-leading terms of the
non-abelian expansion to the second-, third- and higher-derivative
corrections to the non-abelian tachyon DBI action.

In superstring theory, the non-abelian expansion of a S-matrix
element  can be found  easily \cite{garousi1}.  The S-matrix
elements of odd number of tachyon vertex operators  are zero, and
expansion of a S-matrix element of even number of tachyons can be
found as the following: Both tachyon and massless transverse
scalars transform in the adjoint representation of $U(N)$ group,
hence, they both have identical non-abelian kinetic terms. The
Feynman amplitudes resulting from the kinetic terms are then
similar in both cases. Therefore, the non-abelian expansion of a
S-matrix element of even number of tachyons and the non-abelian
expansion of the S-matrix element in which the tachyons are
replaced by the transvers scalars should be similar. On the other
hand,  the non-abelian expansion of  a S-matrix element of even
number of massless scalars is known, \ie sending all Mandelstam
variables to zero. Using similar steps for the S-matrix element of
tachyons, one would find the non-abelian expansion of the tachyon
amplitude \cite{garousi1}. In the bosonic theory, however, the
S-matrix elements of odd number of tachyons are non-zero and the
S-matrix elements in which the tachyons are replaced by massless
scalar fields are zero. Hence, one can not use the above
prescription to find the non-abelian expansion of the S-matrix
element of odd number of tachyons.

 An observation made in \cite{garousi2} is that
the S-matrix elements of four   tachyons and the S-matrix element
in which the tachyons are replaced by the massless scalars can be
written in a universal form.  We speculate  that this property
holds for any S-matrix element of even number of tachyons. This
indicates that the on-shell mass of tachyons/scalars does not
appear in the S-matrix element in the universal from.
In order to reproduce the leading terms of the expansion by a
tachyon action, however, one must use the relations in which  the
Mandelstam variables are satisfied, \eg \reef{cons1}. Using the
fact that these relations  involve the mass of tachyon, one
realizes that the tachyon action  should include the tachyon mass.
One may take the fact that  the tachyon mass does not appear in
the string theory side as an evidence  that the mass in field
theory side is arbitrary. However, this field theory is valid only
for those open string vertex operators whose S-matrix elements can
be written in the universal form, \ie the open string tachyon of N
non-BPS D$_p$-branes\footnote{In \cite{garousi}, it was shown that
the S-matrix element of four massive scalars with vertex operator
$\l\int dx (\z_i\prt^nX^i)e^{ik\cdot X}$ where
$k^2=-(n-1)/\alpha'$, can be written in the universal form. Then
it was concluded that various  coupling of these scalars are given
by the tachyon DBI action in which the mass of tachyon in the
potential $V(T)=e^{\pi\alpha'm^2T^2}$ is replaced by
$m^2=(n-1)/\alpha'$. However, the above vertex is not a primary
operator. The primary massive vertex operators can not be written
in the universal form. Hence, the tachyon DBI action can be used
only for
 the tachyon vertex operator of N non-BPS D$_p$-branes.}.


In this paper, we would like to calculate  the S-matrix element of
four tachyons and one gauge field, and find the non-abelian
expansion of the amplitude. One  can find this expansion in two
different ways. 1-Comparing the amplitude with the amplitude of
four transverse scalars and one gauge field. Non-abeian expansion
of the latter  amplitude  can be found by sending all Mandelstam
variables to zero. Similar expansion for the S-matrix element of
tachyons gives the non-abelian expansion. 2-Using the on-shell
constraints that the Mandelstam variables satisfy, \ie
\reef{cons1}, one may write  the S-matrix element of tachyons in
the universal form. Then send all Mandelstam variables to zero. In
the present paper, we will use the first approach to find the
non-abelian expansion. In sect.2, we will calculate the S-matrix
element of four tachyons and one gauge field.  In order to find
the non-abelian expansion of the S-matrix element, we shall
calculate the S-matrix element of four scalars and one gauge field
in sect.3.1, and then find the non-abelian expansion of the
tachyon amplitude. In sect.4, we shall show that while the first
leading order terms of the expansion which are tachyon and
massless poles are reproduced by non-abelian kinetic term of
tachyon, the next leading order terms which are tachyon poles,
massless poles and contact terms are fully consistent with
non-abelian tachyon DBI action. This calculation fixes the
four-tachyon coupling in the tachyon DBI action without on-shell
ambiguity.

\section{Tachyon amplitude in superstring theory}

Using the world-sheet conformal field theory technique \cite{jp},
one can evaluate a 5-point function by evaluating the correlation
function of their corresponding vertex operators. Performing the
correlations, one finds  that the integrand has $SL(2,R)$
symmetry. Then one should fix this symmetry by fixing position of
three vertices in the real line. Different fixing of these
positions give different ordering of the five vertices in the
boundary of the world-sheet. One should add all non-cyclic
permutation of the vertices to get the correct scattering
amplitude. For the purpose of comparing string theory S-matrix
element with field theory S-matrix element, it is enough to
consider only a special ordering, \ie $(x_1=0, x_2, x_3,
x_4=1,x_5=\infty)$ where the label 5 refers to the gauge field in
the amplitude. With this ordering, one finds that the S-matrix
element has the Chan-Paton factor $\Tr(\l^1\l^2\l^3\l^4\l^5)$.
After  fixing the $SL(2,R)$, one ends up with  one double integral
which can be performed. The result is a multiple of  two Beta
functions and one Hypergeometric function \cite{YK}.

In particular, the S-matrix element of four tachyons and one gauge
field of N non-BPS D$_p$-branes in superstring theory is given by
the following correlation function: \beqa {\cal A}^{TTTTA}&\sim&
\int dx_1dx_2dx_3dx_4dx_5~\langle V_0^{T}V_0^{T}V_0^{T}V_{-1}^{T}
V_{-1}^{A}\rangle \labell{amp0}\eeqa where the vertex operators
for tachyons and gauge field are given by
 \beqa V_0^{T}&=&\z^{\alpha}(2ik\inn \psi)e^{2ik\inn X}\Lambda^{\alpha} \nonumber\\
 V_{-1}^{T}&=&\z^{\alpha}e^{2ik\inn X}e^{-\Phi} \Lambda^{\alpha} \nonumber\\
  V_{-1}^{A}&=&\xi_a^{\alpha}\psi^ae^{2ik\inn X}e^{-\Phi} \Lambda^{\alpha}\nonumber \eeqa
where   $\alpha=1,2,\cdots,N^2$ and $N$ is the number of
$D_p$-branes. $\z$ is polarization of tachyon which specifies one
of the $N^2$ open string tachyons existed in the $D_p$-branes
system. Similarly, $\xi_a^{\alpha}$ is the polarization of gauge
field where the index $a$ specifies the component of gauge field
in the world volume of $D_p$-branes. The matrices
$\Lambda^{\alpha}$ are the complete bases in terms of which
 the Chan-Paton matrix can be expanded, \ie
 $\l=\z^{\alpha}\L^{\alpha}$ for tachyon,
 and $\xi_a\l=\xi_a^{\alpha}\L^{\alpha}$ for gauge field.
In the above vertices,  $k$ is the world volume momentum of the
open string states. The on-shell condition for tachyon is
$k^2=1/(2\alpha')$, and for gauge field is $\xi_a k^a=0$ and
$k^2=0$. $X^a$, $\psi^a$ and $\Phi$ are the usual world-sheet
fields of  the fermionic string (see \cite{jp} for details). The
two dimensional world sheet fields have the standard propagators:
 \beqa
\langle X^{\mu}(z)\,X^{\nu}(w)\rangle&=&-\eta^{\mu \nu}\,\log(z-w)
\nonumber\\
\langle\psi^{\mu}(z)\,\psi^{\nu}(w)\rangle&=&-\frac{\eta^{\mu
\nu}}{z-w} \nonumber\\
\langle\Phi(z)\,\Phi(w)\rangle&=&-\log(z-w) \labell{standard}
 \eeqa
Using the conformal field theory technique, one can evaluate the
correlation functions and  show that integrand has the $SL(2,R)$
symmetry. Fixing this symmetry as $(x_1=0,\,x_2,\,x_3,\,
x_4=1,\,x_5=\infty)$, one finds the final result as

 \beqa {\cal A}^{TTTTA} &=&-4i\alpha'T_p
\Tr(\lambda^1\lambda^2\lambda^3\lambda^4\lambda^5)\nonumber\\&&\times
\left(\frac{}{}(-s_{23}-1)a'k_1\cdot \xi_5+(s_{13}+1)b'k_2\cdot
\xi_5+(-s_{12}-1)c'k_3\cdot \xi_5\right)\labell{amp}\eeqa where
$a',~b'$, and $~c'$ are the double integrals that left over after
fixing the position of three vertex operators. These integrals are
discussed in the appendix A. We have also normalized the amplitude
by $-4i\alpha'T_p$. To compare the above S-matrix element with the
corresponding amplitude in field theory, one needs only  to keep
the first term, \ie $(a'k_1\cdot \xi_5)$ term. In terms of the
Beta  and Hypergeometric functions, this term is
 \beqa {\cal A}^{TTTTA}&\!\!\!\!=\!\!\!\!&-4i\alpha'T_p\Tr
(\l^1\l^2\l^3\l^4\l^5)k_1\cdot\xi_5\nonumber\\
&&\left(\frac{}{}(-s_{23}-1)\beta(-s_{12},-s_{23}-1)
 \beta(-s_{45}-\frac{1}{2},-s_{34})\right.\labell{bbf}\\&&\left.\times
{}_{3}F_2(-s_{12},-s_{45}-\frac{1}{2},s_{15}-s_{23}-s_{34}-\frac{1}{2};
-s_{12}-s_{23}-1,-s_{34}-s_{45}-\frac{1}{2};1)\right)\nonumber\eeqa
where the definition of kinematic factors $s_{ij}$ is \beqa
  s_{ij}=-\alpha'(k_i+k_j)^2\labell{kinematic}\eeqa
The number of independent kinematic factors in the scattering
amplitude of $n$ states is $\frac{n}{2}(n-3)$ \cite{ZK}. In the
present case, there are 5 independent kinematic factors. In
writing the above amplitude, we have chosen them  as
$s_{12},s_{23},s_{34},s_{45},s_{15}$. Using  conservation of
momentums and the on-shell conditions
$k_1^2=k_2^2=k_3^2=k_4^2=1/(2\alpha'),\,k_5^2=0$, one finds that
the other kinematic factors $s_{13},s_{14},s_{24},s_{25},s_{35}$
can be written in terms of independent ones as \beqa
s_{13}&=&s_{45}-s_{12}-s_{23}-\frac{3}{2}\nonumber\\s_{14}&=&s_{23}-s_{15}-s_{45}-1
\nonumber\\s_{24}&=&s_{15}-s_{23}-s_{34}-\frac{3}{2}
\labell{cons1}\\s_{25}&=&s_{34}-s_{12}-s_{15}-1\nonumber\\s_{35}&=&s_{12}-s_{45}-s_{34}-1\nonumber\eeqa
One can also show that the Mandelstam variables satisfied the
following relation: \beqa
\sum_{i<j}s_{ij}&=&-6\labell{onshell}\eeqa For later purposes , it
is convenient to absorb the factor $(s_{23}+1)$ in \reef{bbf} into
the first Beta function, \ie  \beqa {\cal
A}^{TTTTA}&\!\!\!\!=\!\!\!\!&-4i\alpha'T_p\Tr
(\l^1\l^2\l^3\l^4\l^5)k_1\cdot\xi_5\nonumber\\
&&\left(\frac{\Gamma(-s_{12})\Gamma(-s_{23})}{\Gamma(-1-s_{12}-s_{23})}
 \beta(-s_{45}-\frac{1}{2},-s_{34})\right.\labell{tachyonamp}\\&&\left.\times
{}_{3}F_2(-s_{12},-s_{45}-\frac{1}{2},s_{15}-s_{23}-s_{34}-\frac{1}{2};
-s_{12}-s_{23}-1,-s_{34}-s_{45}-\frac{1}{2};1)\right)\nonumber\eeqa
From  poles of the Beta function in \reef{bbf}, one realizes that
the amplitude has  tachyon, massless and infinite tower of massive
poles. Some of the  tachyon and massless  poles should be
reproduced, in field theory, by non-abelian kinetic term. It is
completely nontrivial to find an expansion for Beta and
Hypergeometric functions  which keeps only the tachyon and
massless poles which are reproduced by non-abelian kinetic terms
and decouples all other poles. Obviously, because of the tachyon
pole, this would not be the usual $\alpha'$ expansion in which
tachyon and massive modes are decoupled. However, as we argued in
the introduction section, this expansion may lead to an effective
action for tachyon and massless fields when second and higher
derivative of tachyon is small in compare with the string scale.
In the next section we shall find such expansion by comparing it
with the S-matrix element of four transverse scalars and one gauge
field.

\section{Non-abelian Expansion of Tachyon Amplitude}

Using the principle that the tachyon action should have the $U(N)$
symmetry, one can find a natural  expansion for the tachyon
amplitude, \ie an expansion whose first leading order terms are
reproduced by non-abelian kinetic terms.  To find this expansion,
we note that the tachyon and transverse scalar field of
D$_8$-brane transform as the adjoint representation of $U(N)$
group, hence, they both have identical non-abelian kinetic term.
The Feynman diagrams resulting from the kinetic term are then
identical, \ie the Feynman diagrams of scalar field can be
converted to the Feynman diagram of tachyon by replacing each
scalar line with the tachyon line (see Fig.1). The Feynman
amplitudes are also "similar", \ie replacing the propagator of
scalars with the propagator of tachyon, one can convert the scalar
amplitude to the tachyon amplitude. On the other hand,  the
natural  expansion of  the S-matrix element of the scalars whose
first leading order terms are reproduced by non-abelian kinetic
term is known, \ie sending all Mandelstam variables to zero. Using
"similar" expansion for the tachyon amplitude, one can find the
desired non-abelian expansion for tachyon amplitudes.  Hence, to
find the non-abelian expansion of the tachyon amplitude,  we need
to recall the non-abelian/$\alpha'$ expansion of the S-matrix
element of four scalars and one gauge field.

\subsection{Massless amplitude}

The S-matrix element of four transverse scalars and one vector of
$N$ $D_8$-beanes  in the supersting theory can be read from the
S-matrix element of five gauge fields \cite{YK, medina}. However,
our notation for open string momenta and $s_{ij}$ is different
from those references. In our notation, the amplitude is given by
\beqa {\cal A}^{\phi\phi\phi\phi A}&\sim& \int
dx_1dx_2dx_3dx_4dx_5~\langle
V_0^{\phi}V_0^{\phi}V_0^{\phi}V_{-1}^{\phi} V_{-1}^{A}\rangle
\labell{amp00}\eeqa where the vertex operators are given as
 \beqa V_0^{\phi}&=&(\partial X^9+2ik\inn \psi~\psi^9)e^{2ik\inn X}\lambda \nonumber\\
 V_{-1}^{\phi}&=&\psi^9 e^{2ik\inn X}e^{-\Phi} \lambda \nonumber\\
  V_{-1}^{A}&=&\xi_a\psi^ae^{2ik\inn X}e^{-\Phi} \lambda \nonumber\eeqa
where $\xi^a$ is the polarization of  gauge field. Performing the
correlators in \reef{amp00}, one finds the following final result:
\beqa {\cal A}^{\phi\phi\phi\phi A}&\!\!\!=\!\!\!&-4i\alpha'T_8
\Tr(\lambda^1\lambda^2\lambda^3\lambda^4\lambda^5
)\left[\frac{}{}k_1\inn\xi_5\left(-\frac{}{}s_{23}~a_1+s_{23}~a_2
-(s_{23}+1)~a_3\right)\right.\labell{amp01}\\&&\left.+
k_2\inn\xi_5\left(\frac{}{}s_{13}~b_1-(s_{13}+1)~b_2
+s_{13}~b_3\right)+
k_3\inn\xi_5\left(-\frac{}{}(s_{12}+1)~c_1+s_{12}~c_2
-s_{12}~c_3\right)\frac{}{}\right]\nonumber
\eeqa where  the coefficients $a_i,b_i,c_i$ for $i=1,2,3$, which
are the double integrals are discussed in the appendix A.  We have
also normalized the amplitude by the factor $-4i\alpha' T_8$. The
definition of the kinematic factors $s_{ij}$ is the same as
before, \ie \reef{kinematic}. However, because of the on-shell
condition $k_1^2=k_2^2=k_3^2=k_4^2=k_5^2=0$ in the present case,
one finds the following relation between $s_{ij}$:\beqa
s_{13}=s_{45}-s_{12}-s_{23}\nonumber\\s_{14}=s_{23}-s_{15}-s_{45}
\nonumber\\s_{24}=s_{15}-s_{23}-s_{34}
\nonumber\\s_{25}=s_{34}-s_{12}-s_{15}\nonumber\\s_{35}=s_{12}-s_{45}-s_{34}\nonumber\eeqa
and $\sum_{i<j}s_{ij}=0$. Again, we  consider only $k_1\cdot\xi_5$
terms. Using the above relations, one can  arrange the  result as
 \beqa {\cal A}^{\phi\phi\phi\phi A}&=&-4i\alpha'T_8\Tr
(\l^1\l^2\l^3\l^4\l^5)k_1\inn\xi_5\nonumber\\&&
\left(-\frac{}{}s_{23}\beta(-s_{12},-s_{23})
\beta(-s_{45},-s_{34}) \right.\nonumber\\&&\times
{}_{3}F_2(-s_{12},-s_{45},s_{15}-s_{23}-s_{34};-s_{12}-s_{23},-s_{34}-s_{45};1)
\nonumber\\&& +s_{23}\beta(1-s_{12},-s_{23})
\beta(-s_{45},1-s_{34})\nonumber\\&&\times
{}_{3}F_2(1-s_{12},-s_{45},1+s_{15}-s_{23}-s_{34};1-s_{12}-s_{23},1-s_{34}-s_{45};1)\nonumber\\&&
-(s_{23}+1)\beta(1-s_{12},-1-s_{23})
\beta(-s_{45},1-s_{34})\nonumber\\&&\left.\times
{}_{3}F_2(1-s_{12},-s_{45},s_{15}-s_{23}-s_{34};-s_{12}-s_{23},1-s_{34}-s_{45};1)\frac{}{}\right)\nonumber
\eeqa   One can $\alpha'$ expand the Beta  and Hypergeometric
functions (see  appendix A) to find the non-abelian/$\alpha'$
expansion of the above scattering amplitude, and compare the
results with the low energy field theory. We have done this
calculation and found that the leading order terms of the
expansion are in perfect agreement with the non-abelian DBI action
\cite{myers}. Similar calculation for the S-matrix element of five
gauge fields has been done in \cite{YK, medina}. However, our main
purpose behind calculating the above scalar amplitude is to find a
guide for expanding the Beta and Hypergeometric functions in the
tachyon amplitude \reef{bbf}. To this end, we first rewrite the
above amplitude as
 \beqa {\cal A}^{\phi\phi\phi\phi A}&\!\!\!\!\!=\!\!\!\!\!&-4i\alpha'T_8\Tr
(\l^1\l^2\l^3\l^4\l^5)k_1\inn\xi_5\nonumber\\&&
\left(\frac{\Gamma(-s_{12})\Gamma(1-s_{23})}{\Gamma(-s_{12}-s_{23})}
\beta(-s_{45},-s_{34}) \right.\nonumber\\&&\times
{}_{3}F_2(-s_{12},-s_{45},s_{15}-s_{23}-s_{34};-s_{12}-s_{23},-s_{34}-s_{45};1)
\nonumber\\&&-
\frac{\Gamma(1-s_{12})\Gamma(1-s_{23})}{\Gamma(1-s_{12}-s_{23})}
\beta(-s_{45},1-s_{34})\nonumber\\&&\times
{}_{3}F_2(1-s_{12},-s_{45},1+s_{15}-s_{23}-s_{34};1-s_{12}-s_{23},1-s_{34}-s_{45};1)\nonumber\\&&
+\frac{\Gamma(1-s_{12})\Gamma(-s_{23})}{\Gamma(-s_{12}-s_{23})}
\beta(-s_{45},1-s_{34})\nonumber\\&&\left.\times
{}_{3}F_2(1-s_{12},-s_{45},s_{15}-s_{23}-s_{34};-s_{12}-s_{23},1-s_{34}-s_{45};1)\frac{}{}\right)\nonumber
\eeqa Note that each term in the above amplitude  has the same
structure as the tachyon amplitude \reef{tachyonamp}. The
non-abelian expansion of the above scalar amplitude is obviously
expansion around $s_{ij}\rightarrow 0$. This sends the argument of
the above functions to different points, \ie the Gamma, Beta and
Hypergeometric functions in the first, second, and third term must
be expanded around, respectively,\beqa
{\rm first\, term}&\rightarrow&\frac{\Gamma(0)\Gamma(1)}{\Gamma(0)}\beta(0,0){}_{3}F_{2}(0,0,0;0,0;1)\nonumber\\
{\rm second \,term}&\rightarrow&-\frac{\Gamma(1)\Gamma(1)}{\Gamma(1)}\beta(0,1){}_{3}F_{2}(1,0,1;1,1;1)\nonumber\\
{\rm third\,
term}&\rightarrow&\frac{\Gamma(1)\Gamma(0)}{\Gamma(0)}\beta(0,1){}_{3}F_{2}(1,0,0;0,1;1)\labell{expanda}\eeqa
Our claim is that to produce the non-abelian expansion for the
tachyon amplitude \reef{tachyonamp}, one should expand the
amplitude around the same points, \ie the arguments of  Gamma,
Beta and Hypergeometric functions in the tachyon amplitude
\reef{tachyonamp} should  be sent  to  the same points. For
example, in order to send the arguments of  Gamma, Beta and
Hypergeometric functions in \reef{tachyonamp} to
$\frac{\Gamma(0)\Gamma(1)}{\Gamma(0)}\beta(0,0){}_{3}F_{2}(0,0,0;0,0;1)$,
one should send
$s_{23}\rightarrow-1,~s_{45},s_{15}\rightarrow-\frac{1}{2},~s_{12},s_{34}\rightarrow0$.
Similarly for other points. Hence, our proposal for non-abelian
expansion of the tachyon amplitude is the following:\beqa {\cal
A}^{TTTTA}&=&\left(\lim
_{s_{23}\rightarrow-1,~s_{45},s_{15}\rightarrow-\frac{1}{2},~s_{12},s_{34}\rightarrow0}{\cal
A}^{TTTTA}\right)
\labell{limit}\\
&&-\left(\lim _{s_{45},s_{15}\rightarrow -\frac{1}{2},~
s_{12},s_{23},s_{34}\rightarrow -1}{\cal A}^{TTTTA}\right) +\left(
\lim _{s_{45},s_{15}\rightarrow -\frac{1}{2},~
s_{12},s_{34}\rightarrow -1,~s_{23}\rightarrow0 }{\cal
A}^{TTTTA}\right)\nonumber\eeqa  Note that if one ignores the
limit signs, the right hand side becomes ${\cal A}^{TTTTA}$.
Moreover, as we already mentioned around eq.\reef{cons1},  we have
chosen the independent Mandelstam variables to be $s_{12}, s_{23},
s_{34}, s_{45}, s_{15}$, hence, the  above is a linear combination
of three  on-shell limits. Now, one can expand the Beta and
Hypergeometric functions in \reef{bbf} around the above points
(see appendix A). The result is the following: \beqa {\cal
A}^{TTTTA} &\!\!\!\!=\!\!\!\!&4i\alpha'T_p
Tr(\lambda^1\lambda^2\lambda^3\lambda^4\lambda^5)\xi_5 \cdot
k_1\labell{Atttta}\\&& \left[
  \left( \frac{s_{23}}{(s_{45}+\frac{1}{2}) s_{12}}
    + \frac{s_{23}}{(s_{15}+\frac{1}{2}) s_{34}}
 + \frac{s_{23}}{s_{12} s_{34}}+ \frac{s_{12}}{s_{23} (s_{45}+\frac{1}{2})}
  + \frac{s_{34}}{s_{23}(s_{15}+\frac{1}{2})}\right.\right.\nonumber\\&&
+\frac{1}{(s_{45}+\frac{1}{2}) s_{12}} +
\frac{1}{(s_{15}+\frac{1}{2}) s_{34}} + \frac{1}{s_{12} s_{34}}+
\frac{1}{s_{23} (s_{45}+\frac{1}{2})}
 + \frac{1}{s_{23}(s_{15}+\frac{1}{2})} \nonumber\\&&
 +\left. \frac{1}{(s_{15}+\frac{1}{2})}+\frac{1}{(s_{45}+\frac{1}{2})}-\frac{1}{s_{23}}
\right) \nonumber\\&&
-\zeta(2)\left(\frac{s_{12}^2}{(s_{45}+\frac{1}{2})}+\frac{s_{23}^2}{(s_{45}+\frac{1}{2})}+
\frac{s_{12}s_{23}}{(s_{45}+\frac{1}{2})}+\frac{2s_{12}}{(s_{45}+\frac{1}{2})}
+\frac{2s_{23}}{(s_{45}+\frac{1}{2})}+\frac{1}{(s_{45}+\frac{1}{2})}\right.\nonumber\\&&
+\frac{s_{23}^2}{(s_{15}+\frac{1}{2})}+\frac{s_{34}^2}{(s_{15}+\frac{1}{2})}+
\frac{s_{23}s_{34}}{(s_{15}+\frac{1}{2})}+
\frac{2s_{23}}{(s_{15}+\frac{1}{2})}+\frac{2s_{34}}{(s_{15}+\frac{1}{2})}+\frac{1}{(s_{1
5}+\frac{1}{2})}\nonumber\\&&
+\frac{s_{23}s_{12}}{s_{34}}+\frac{s_{23}(s_{15}+\frac{1}{2})}{s_{34}}+\frac{s_{12}}{s_{34}}
+\frac{(s_{15}+\frac{1}{2})}{s_{34}}
+\frac{s_{23}s_{34}}{s_{12}}+\frac{s_{23}(s_{45}+\frac{1}{2})}{s_{12}}+\frac{s_{34}}{s_{12}}
\nonumber\\&&+\frac{(s_{45}+\frac{1}{2})}{s_{12}}+\frac
{s_{12}(s_{15}+\frac{1}{2})}{s_{23}}+\frac
 {s_{34}(s_{45}+\frac{1}{2})}{s_{23}} +\frac
{(s_{45}+\frac{1}{2})}{s_{23}}-\frac{(s_{15}+\frac{1}{2})(s_{45}+\frac{1}{2})}{s_{23}}\nonumber\\&&
\left.\left.+\frac{(s_{15}+\frac{1}{2})}{s_{23}}
 -s_{12}-s_{34}+s_{15}+s_{45}-3s_{23}-3\right)+\cdots \right]\nonumber\eeqa
where dots represent terms that are proportional to $\zeta(3)$,
$\zeta(4)$, and so on. This is the non-abelian expansion of the
tachyon amplitude \reef{bbf} we were after.  As we anticipated
before, the above non-abelian expansion keeps some of the tachyon
and massless poles of amplitude \reef{bbf} and decouples all other
poles. Unlike the non-abelian expansion of the S-matrix elements
of massless states, the above expansion is not an $\alpha'$
expansion. It has the same $\z(n)$ expansion as  the massless
scalar amplitude. However, unlike the scalar amplitude, the
different terms at each $\z(n)$ have different $\alpha'$ order. In
the field theory side, this indicates that the terms at each
$\z(n)$ may have different $\alpha'$ order terms. In the next
section, we shall show that the terms in the first three lines
above are reproduced by non-abelian kinetic terms \ie the terms in
the first line of \reef{expandL}, and the terms proportional to
$\z(2)$ are reproduced by all terms in \reef{expandL} which have
obviously different $\alpha'$ order.
\section{Amplitude in Effective Field Theory}
The states that appear as on-shell or off-shell in the above
S-matrix element are only tachyon and gauge field. We are
interested then in the part of effective field theory of
$D_p$-branes which includes only tachyon and gauge fields. Hence,
we are looking for the effective action of $D_9$-brane in flat
background. The proposal is  the non-abelian tachyon DBI action of
$D_9$-branes which is given by\footnote{ One may use the
non-abelin DBI action of two non-BPS D-branes to find the
effective action of D-brane-anti-D-brane \cite{mohammad}.} \beqa
{\cal L}^{\rm BI}&=&-T_9\,{\rm STr}\left(V(T)\frac{}{}
\sqrt{-\det(\eta_{ab}+2\pi\alpha' F_{ab}+2\pi\alpha'
D_aTD_bT)}\right)\,\,,\labell{qt1}\eeqa where the $T_9$ is the
brane tension. The trace on the non-abelian matrices is the
symmetric trace. The non-abelian field strength and covariant
derivative of tachyon are, respectively,
 \beqa F^{ab}=
\partial^aA^b-\partial^bA^a-i[A^a,A^b],~~~~
D_aT=\partial_aT-i[A_a,T] \nonumber\eeqa  The tachyon potential
has the following expansion: \beqa
V(T)&=&1+\pi\alpha'm^2T^2+\frac{1}{2}(\pi\alpha'm^2T^2)^2+\cdots\labell{potential}
\eeqa  where $m^2$ is the mass squared of tachyon, \ie
$m^2=-1/(2\alpha')$. The above expansion is consistent with the
potential $V(T)=e^{\pi\alpha'm^2T^2}$ which is the tachyon
potential of BSFT \cite{DK}\footnote{Expanding the square root,
and normalizing tachyon as $\frac{1}{2}\pi T^2\rightarrow 2T^2$,
one finds $S=-T_9\,e^{-T^2}\left(1+4\alpha'D_aTD^aT+\cdots\right)$
which is the two derivative truncation of BSFT action proposed in
\cite{JAM}.}.

 In the non-abelian  field theory \reef{qt1}, $A_a$ and $T$ are in the adjoint
representation of the gauge symmetry $U(N)$, where $N$ is the
number of $D_9$-branes. That is $A_a=A_a^{\alpha}\Lambda^{\alpha}$
and $T=T^{\alpha}\L^{\alpha}$ where the  hermitian matrices
$\Lambda^{\alpha}$ are the adjoint representation  of the $U(N)$
group. Our conventions for $\Lambda^{\alpha}$ are  \beqa
\sum_\alpha
\L^{\alpha}_{ij}\L^{\alpha}_{kl}=\delta_{il}\delta_{jk}\,\,,\,\,~~~
\Tr(\Lambda^{\alpha}\Lambda^{\beta})&\!\!\!=\!\!\!&\delta^{\alpha\beta},
~~~[\L^{\alpha},\L^{\beta}]=if^{\alpha\beta\gamma}\L^\gamma\nonumber\eeqa
One can express  the structure constant in terms of the group
generators \beqa
if^{\alpha\beta\gamma}=\Tr([\L^{\alpha},\L^{\beta}]\L^{\gamma})\nonumber\eeqa
Using  the following expansion, one can expand the square root in
the non-abelian action \reef{qt1} to  produce various interacting
terms \beqa \sqrt{ -\det(M_0+M)}&=&\sqrt{
-\det(M_0)}\left(1+\frac{1}{2}Tr\left(M_{0}^{-1}M\right)-\frac{1}{4}Tr\left(M_{0}^{-1}MM_{0}^{-1}M\right)
\right.\nonumber\\
&&\left.+\frac{1}{8}\left(Tr\left(M_{0}^{-1}M\right)\right)^2+\frac{1}{6}Tr\left(M_{0}^{-1}MM_{0}^{-1}MM_{0}^{-1}M\right)
\right.\nonumber\\
&&\left.-\frac{1}{8}\left(Tr\left(M_{0}^{-1}M\right)\right)^3+\cdots\,\right)\labell{q13}\eeqa
In \reef{qt1},  $M_0$ and $M$ are \beqa
M_0&=&\eta_{ab}\nonumber\\M&=&2\pi\alpha'( F_{ab}+ D_aTD_bT)
\nonumber\eeqa The terms of the above expansion which has
contribution to the S-matrix element of one gauge field and four
tachyons are the following: \beqa {\cal
L}&=&-T_9(\pi\alpha')\Tr\left(m^2T^2+D_aTD^aT-(\pi\alpha')F_{ab}F^{ba}\right)\labell{expandL}\\
&&-T_9(2\pi\alpha')^2S\Tr\left(\frac{m^4}{8}T^4+\frac{m^2}{4}T^2D_aTD^aT-\frac{1}{8}(D_aTD^aT)^2\right)\nonumber\\
&&-T_9\frac{(2\pi\alpha')^3}{2}S\Tr\left(D^aTD_bTF^{bc}F_{ca}
-\frac{1}{4}D^aTD_aTF^{bc}F_{cb}-\frac{m^2}{4}TTF^{ab}F_{ba}\right)\nonumber\eeqa
Note that the above terms are not ordered in terms of power of
$\alpha'$. However, we shall show that the terms in the first line
which we call them kinetic order terms, reproduce the first
leading terms in \reef{Atttta}, \ie the terms  which are not
proportional to $\z(n)$, and the terms in the second and third
lines in combination with the terms in the first line reproduce
the next leading order terms in \reef{Atttta}, \ie the terms
which are proportional to $\z(2)$.
\subsection{Kinetic  order terms}

The non-abelian kinetic terms in the first line of \reef{expandL}
give
 the following  gauge
field and tachyon propagators, respectively, \beqa G_{\alpha
\beta}^{ab}(A) &=&\frac{i\delta^{ab}\delta_{\alpha
\beta}}{(2\pi\alpha')^2
T_9\left(-k^2\right)}\nonumber\\
G_{\alpha \beta}(T)&=&\frac{i\delta_{\alpha \beta}}{(2\pi\alpha')
T_9(-k^2-m^2)}\labell{propa}
 \eeqa where $\alpha, \beta $
are the group indices and $a,b$ are the world volume indices, and
various three and four field couplings. From these couplings, one
finds three Feynman diagrams contributing to the S-matrix element
of one gauge field and four tachyons (see fig.1). In these
diagrams the wavy lines represent gauge field, and straight lines
represent tachyon fields. The label 5 is for gauge field and
$1,2,3,4$ for tachyons. We will choose  special ordering for
non-abeilian fields such that each diagram produces
$\Tr(\l^1\l^2\l^3\l^4\l^5)$ factor for its corresponding
amplitude, the same group factor that appears in string theory
amplitude \reef{Atttta}.  Our convention for momenta is so that
all momenta in each vertex are inwards. Now we compute the Feynman
amplitude corresponding to these graphs. We write the amplitude as
\beqa {\cal A}_k^{TTTTA}&=&A_k+A'_k+A''_k\nonumber\eeqa where $
A_k$ corresponds to diagram (a), $A'_k$ corresponds to diagram
(b), and $A''_k$ corresponds to diagram (c).
\begin{figure}
{ \epsfbox {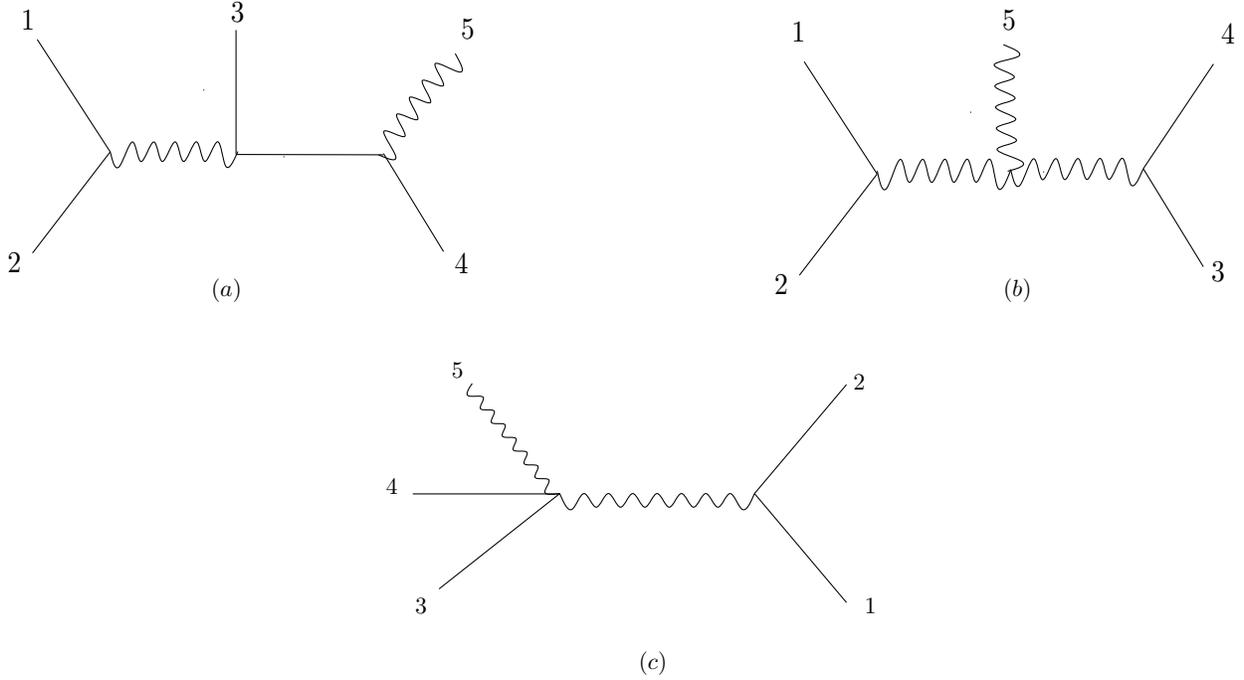}}
 \caption{ Feynman diagrams resulting from kinetic terms.  }
\end{figure}

  Using Feynman rules, one can write $ A_k(T_1T_2T_3T_4A_5)$ as the following:
  \beqa A_k(T_1T_2T_3T_4A_5)&=&
V^a_{\alpha}(T_1T_2A)~ G^{ab}_{\alpha\beta}(A)~
V^{b}_{\beta\gamma}(AT_3T)~ G_{\gamma\l}(T)~
V_{\l}(TT_4A_5)\labell{amp1}\eeqa where the propagators are given
in \reef{propa}. The  vertices can be written in terms of
structure constant or in terms of trace of $\l$'s. In this
section, we use the standard notation and write the vertices of
kinetic terms  in terms of structure constant. Hence, the above
vertices are \beqa
V^a_{\alpha}(T_1T_2A)&=&(2\pi\alpha'T_9)\z_1^{\alpha_1}\z_2^{\alpha_2}
(k_1^a-k_2^a)f^{\alpha_1\alpha_2\alpha}\nonumber
\\ V^b_{\beta\gamma}(AT_3T)&=&
(2\pi\alpha'T_9)\z_3^{\alpha_3}\left(k^b-k_3^b\right)f^{\beta\alpha_3\gamma}\nonumber \\
V_{\l}(TT_4A_5)&=&(2\pi\alpha'T_9)\z_4^{\alpha_4}(k-k_4)\cdot\xi_5^{\alpha_5}f^{\l\alpha_4\alpha_5}\labell{vertex}\eeqa
where $k$ is momentum of off-shell tachyon (see fig.2). Replacing
propagators and vertices in \reef{amp1}, one finds
\begin{figure}
{ \epsfbox {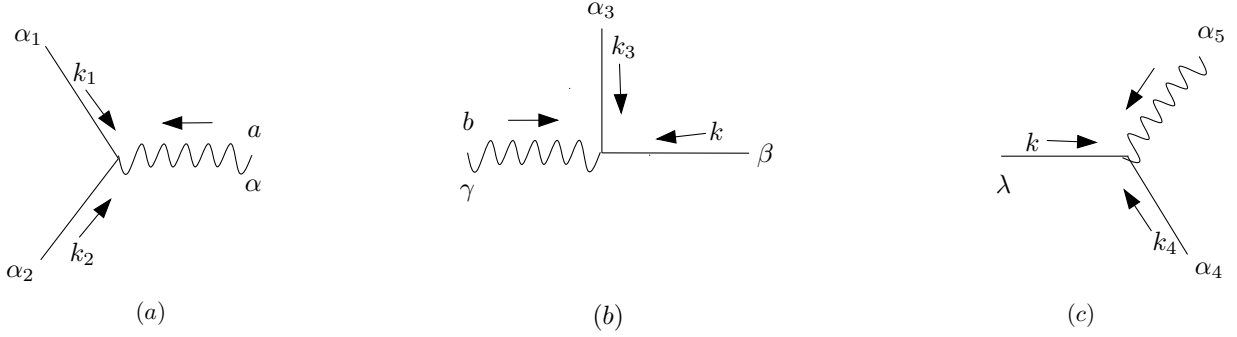}} \caption{ Vertices of two tachyons and
one gauge field. a)Two tachyons are on-shell. b)One tachyon is
on-shell. c)One tachyon and the gauge field are on-shell. }
\end{figure}
\beqa A_k(T_1T_2T_3T_4A_5)&\!\!\!=\!\!\!&
2T_9\z_1^{\alpha_1}\z_2^{\alpha_2}\z_3^{\alpha_3}\z_4^{\alpha_4}
k_4\cdot\xi_5^{\alpha_5}\frac{(k_1-k_2)\cdot \left(k_4+k_5-k_3
\right)}{(k_1+k_2)^2((k_4+k_5)^2+m^2)}f^{\alpha_1\alpha_2\alpha}
f^{\alpha\alpha_3\beta}f^{\beta\alpha_4\alpha_5}\nonumber\eeqa
Note that the amplitude is symmetric under changing
$1\leftrightarrow2$. The amplitude in terms of  $s_{ij}$ can be
rewritten as \beqa
A_k(T_1T_2T_3T_4A_5)&=&-(2\alpha'T_9)\z_1^{\alpha_1}
\z_2^{\alpha_2}\z_3^{\alpha_3}\z_4^{\alpha_4}k_4\cdot\xi_5^{\alpha_5}f^{\alpha_1\alpha_2\alpha}
f^{\alpha\alpha_3\beta}f^{\beta\alpha_4\alpha_5}
\nonumber\\&&\times(-\frac{1}{s_{12}}+\frac{1}{s_{45}+\frac{1}{2}}+\frac{2s_{23}}{(s_{45}+\frac{1}{2})s_{12}}
+\frac{2}{(s_{45}+\frac{1}{2})s_{12}}) \nonumber\eeqa The
amplitude in the above form that we shall compare it with
\reef{Atttta}, is not symmetric under changing $1\leftrightarrow
2$, however, using the relations in \reef{cons1}, one may rewrite
it in a symmetric form. The color factor of the diagram (a) in
figure 1 can be expressed in terms of traces of the group
generators as the following \beqa -i
f^{\alpha_1\alpha_2\alpha}f^{\alpha\alpha_3\beta}f^{\beta\alpha_4\alpha_5}&\!\!\!\!=\!\!\!\!&
-\Tr\left([\L^{\alpha_1},\L^{\alpha_2}]\L^{\alpha}\right)
f^{\alpha\alpha_3\beta}f^{\beta\alpha_4\alpha_5}\nonumber\\&=&
\Tr\left([\L^{\alpha_1},\L^{\alpha_2}][\L^{\alpha_3},\L^{\beta}]\right)i
f^{\beta\alpha_4\alpha_5}\nonumber\\&=&
\Tr\left([\L^{\alpha_1},\L^{\alpha_2}][\L^{\alpha_3},[\L^{\alpha_4},\L^{\alpha_5}]]\right)\nonumber
\eeqa In terms of $\l$'s, one can write \beqa -i\z_1^{\alpha_1}
\z_2^{\alpha_2}\z_3^{\alpha_3}\z_4^{\alpha_4}k_4\cdot\xi_5^{\alpha_5}
f^{\alpha_1\alpha_2\alpha}f^{\alpha\alpha_3\beta}f^{\beta\alpha_4\alpha_5}&\!\!\!\!=\!\!\!\!&
k_4\cdot\xi_5\Tr\left([\l^{1},\l^{2}][\l^{3},[\l^{4},\l^{
5}]]\right)\nonumber\eeqa This includes among other things our
favorite ordering that we had chosen in the string theory side,
\ie $\Tr(\l^{1}\l^{2}\l^{3}\l^{4}\l^{5})$. Using the fact that in
the string theory side \reef{amp} there is no term proportional to
$k_4\cdot\xi_5$, one should use conservation of momentum in the
field theory side to rewrite
$k_4\cdot\xi_5=-(k_1+k_2+k_3)\cdot\xi_5$. So the above term does
has contribution to the $k_1\cdot\xi_5$ terms.

 By changing the labels of fields in the Feynman diagram (a),
  one can find  other diagrams that produce the ordering
$\Tr(\l^{1}\l^{2}\l^{3}\l^{4}\l^{5})$. The color factor of
$A_k(T_3T_4T_2T_1A_5)$ is given by \beqa -i\z_1^{\alpha_1}
\z_2^{\alpha_2}\z_3^{\alpha_3}\z_4^{\alpha_4}k_1\cdot\xi_5^{\alpha_5}
f^{\alpha_3\alpha_4\alpha}f^{\alpha\alpha_2\beta}f^{\beta\alpha_1\alpha_5}&=&
k_1\cdot\xi_5\Tr\left([\l^{3},\l^{4}][\l^{2},[\l^{1},\l^{5}]]\right)\nonumber\\
&=&k_1\cdot\xi_5\Tr(\l^1\l^2\l^3\l^4\l^5)+\cdots\nonumber\eeqa
The other part of amplitude  $A_k(T_3T_4T_2T_1A_5)$ can be read
from $A_k(T_1T_2T_3T_4A_5)$ by changing the numbers 1,2,3,4,5 to
the 3,4,2,1,5. That is \beqa
A_k(T_3T_4T_2T_1A_5)&=&-(2i\alpha'T_9)k_1\cdot\xi_5\Tr(\l^1\l^2\l^3\l^4\l^5)\nonumber\\
&&\times
(\frac{1}{s_{34}}-\frac{1}{s_{15}+\frac{1}{2}}-\frac{2s_{23}}{(s_{15}+\frac{1}{2})s_{34}}
-\frac{2}{(s_{15}+\frac{1}{2})s_{34}})+\cdots\nonumber  \eeqa
where dots refer to the terms that have ordering other than
$\Tr(\l^{1}\l^{2}\l^{3}\l^{4}\l^{5})$. Two other ordering of
diagram (a) that produce $\Tr(\l^{1}\l^{2}\l^{3}\l^{4}\l^{5})$ are
$A_k(T_2T_3T_4T_1A_5)$ and $A_k(T_3T_2T_1T_4A_5)$. One  finds
these amplitudes from the original amplitude
$A_k(T_1T_2T_3T_4A_5)$, by changing the numbers 1,2,3,4,5 to
2,3,4,1,5 and to 3,2,1,4,5. The results are the following:\beqa
A_k(T_2T_3T_4T_1A_5)&=&(2i\alpha'T_9)k_1\cdot\xi_5\Tr(\l^1\l^2\l^3\l^4\l^5)\nonumber\\
&&\times(-\frac{1}{s_{23}}+\frac{1}{s_{15}+\frac{1}{2}}+\frac{2s_{34}}{(s_{15}+\frac{1}{2})s_{23}}
+\frac{2}{(s_{15}+\frac{1}{2})s_{23}})+\cdots\nonumber\\
A_k(T_3T_2T_1T_4A_5)&=&-(2i\alpha'T_9)k_4\cdot\xi_5\Tr(\l^1\l^2\l^3\l^4\l^5)\nonumber\\
&&\times(-\frac{1}{s_{23}}
+\frac{1}{s_{45}+\frac{1}{2}}+\frac{2s_{12}}{(s_{45}+\frac{1}{2})s_{23}}
+\frac{2}{(s_{45}+\frac{1}{2})s_{23}})+\cdots\nonumber\eeqa Note
that in amplitude  $A_k(T_3T_2T_1T_4A_5)$ one should use
conservation of momentum to produce the factor $k_1\cdot\xi_5$ in
which we are interested in the string theory side.

 Now consider the
Feynman  diagram (b) in the fig.1.  The amplitude is given by:
\beqa A'_k(T_1T_2A_5T_3T_4)&=&{ V}^a_{\alpha}(T_1T_2A)~
G^{ab}_{\alpha\beta}(A)~{ V}^{bc}_{\beta\gamma}(AA_5A)~
G^{cd}_{\gamma\l}(A)~{ V}^{d}_{\l}(AT_3T_4)\labell{amp2} \eeqa
Where the vertex of one on-shell and two off-shell gauge fields
\begin{figure}
{ \epsfbox {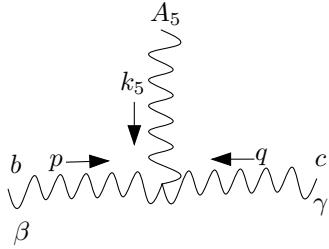}}
 \caption{ Vertex of one on-shell gauge field and two off-shell gauge fields.  }
\end{figure}
is the following \beqa {
 V}^{bc}_{\beta\gamma}(AA_5A)&\!\!\!\!=\!\!\!\!&-T_9(2\pi\alpha')^2f^{\beta\gamma\alpha_5}
\left((k_5-q)^b(\xi_5^c)^{\alpha_5}+(p-k_5)^c(\xi_5^b)^{\alpha_5}+
\eta^{bc}(q-p)\cdot\xi_5^{\alpha_5}\right)\nonumber\eeqa
where $p$ and $q$ are the momentum of off-shell gauge fields (see
 fig.3).  Replacing the above vertex and the other vertex from
\reef{vertex} in \reef{amp2},  one finds  \beqa
A'_k(T_1T_2A_5T_3T_4)&=&-(\alpha'^2T_9)\z_1^{\alpha_1}\z_2^{\alpha_2}\z_3^{\alpha_3}\z_4^{\alpha_4}
f^{\alpha_1\alpha_2\alpha}f^{\alpha\beta\alpha_5}f^{\alpha_3\alpha_4\beta}
\frac{1}{s_{12}s_{34}}\nonumber\\&&
(2(k_1\cdot\xi_5^{\alpha_5}+k_2\cdot\xi_5^{\alpha_5})(k_1-k_2)\cdot(k_3-k_4))\nonumber\\&&
-((k_1\cdot\xi_5^{\alpha_5}-k_2\cdot\xi_5^{\alpha_5})(k_1+k_2-k_5)\cdot(k_3-k_4))\nonumber\\&&
+((k_3\cdot\xi_5^{\alpha_5}-k_4\cdot\xi_5^{\alpha_5})(k_1-k_2)\cdot(k_3+k_4-k_5))\nonumber\eeqa
In the last line above, one should again use
$k_4\cdot\xi_5=-(k_1+k_2+k_3)\cdot\xi_5$. To compare with the
specific terms that we had chosen in string theory side, we
consider only the coefficients $k_1\cdot\xi_5$, \ie  \beqa
A'_k(T_1T_2A_5T_3T_4)
&=&-(\alpha'T_9)\z_1^{\alpha_1}\z_2^{\alpha_2}\z_3^{\alpha_3}\z_4^{\alpha_4}
f^{\alpha_1\alpha_2\alpha}f^{\alpha\alpha_5\beta}f^{\beta\alpha_3\alpha_4}
k_1\cdot\xi_5^{\alpha_5}\nonumber\\
&&\times(\frac{1}{s_{34}}+\frac{1}{s_{12}}+
\frac{4s_{23}}{s_{12}s_{34}}+\frac{4}{s_{12}s_{34}})+\cdots
\nonumber\eeqa where dots refer to the terms that have
coefficients $k_2\cdot\xi_5$ and $k_3\cdot\xi_5$.  The color
factor in the above equation can be expressed as the following
\beqa
-i\z_1^{\alpha_1}\z_2^{\alpha_2}\z_3^{\alpha_3}\z_4^{\alpha_4}
f^{\alpha_1\alpha_2\alpha}f^{\alpha\alpha_5\beta}f^{\beta\alpha_3\alpha_4}k_1\cdot\xi_5^{\alpha_5}&=&
k_1\cdot\xi_5\Tr\left([\l^{1},\l^{2}][\l^{5},[\l^{3},\l^{4}]]\right)\nonumber\eeqa
Obviously  the above factor includes the ordering
$-\Tr(\l^{1}\l^{2}\l^{3}\l^{4}\l^{5})$. By different labelling the
diagram (b), one finds no other distinct diagram that produces the
ordering $\Tr(\l^{1}\l^{2}\l^{3}\l^{4}\l^{5})$.

Finally, we consider the diagram (c) in fig1. The amplitude is
given  as \beqa A''_k(A_5T_3T_4T_1T_2)&=&{ V
}_{\alpha}^a(A_5T_3T_4A)G_{\alpha\beta}^{ab}(A){
 V}_{\beta}^b(AT_1T_2)\labell{amp3}\eeqa
 The vertex of two tachyons and two gauge fields in which only one of the gauge fields is off-shell is given by
 \beqa  {
V}^a_{\alpha}(A_5T_3T_4A)&=&-iT_9(2\pi\alpha')\z_3^{\alpha_3}\z_4^{\alpha_4}
(\xi_5^a)^{\alpha_5}\left(f^{\alpha_5\alpha_3\rho}f^{\rho\alpha\alpha_4}
+f^{\alpha_5\alpha_4\rho}f^{\rho\alpha\alpha_3}\right)
\nonumber\eeqa After replacing the gauge field propagator and
vertices in \reef{amp3}, one finds  \beqa
A''_k(A_5T_3T_4T_1T_2)&=&-(i\alpha'T_9)\z_1^{\alpha_1}\z_2^{\alpha_2}\z_3^{\alpha_3}\z_4^{\alpha_4}
\left(f^{\alpha_5\alpha_3\rho}f^{\rho\alpha\alpha_4}
+f^{\alpha_5\alpha_4\rho}f^{\rho\alpha\alpha_3}\right)if^{\alpha\alpha_1\alpha_2}\nonumber\\&&
\times(k_1\cdot\xi_5^{\alpha_5}-k_2\cdot\xi_5^{\alpha_5})\frac{1}{s_{12}}\nonumber
\eeqa The color factor can be written as \beqa
\z_1^{\alpha_1}\z_2^{\alpha_2}\z_3^{\alpha_3}\z_4^{\alpha_4}\left(f^{\alpha_5\alpha_3\rho}f^{\rho\alpha\alpha_4}
+f^{\alpha_5\alpha_4\rho}f^{\rho\alpha\alpha_3}\right)if^{\alpha\alpha_1\alpha_2}\nonumber\\
=\z_1^{\alpha_1}\z_2^{\alpha_2}\left(\Tr\left([\L^{\alpha_5},\l^{3}][\L^{\alpha},\l^{4}]\right)
+\Tr([\L^{\alpha_5},\l^{4}][\L^{\alpha},\l^{3}])\right)if^{\alpha\alpha_1\alpha_2}\nonumber\\
=Tr([\l^{5},\l^{3}]\left[[\l^{1},\l^{2}],\l^{4}\right])
+\Tr([\l^{5},\l^{4}]\left[[\l^{1},\l^{2}],\l^{3}\right])\nonumber\eeqa
The final result will be\beqa
A''_k(A_5T_3T_4T_1T_2)&=&(i\alpha'T_9)\left(\frac{}{}\Tr(\l^{1}\l^{2}\l^{3}\l^{4}\l^{5})+\cdots\right)
(k_1\cdot\xi_5-k_2\cdot\xi_5)\frac{1}{s_{12}}\nonumber \eeqa where
dots refer to other orderings. By changing the labels in diagram
(c), one finds two other distinct diagrams that produce the
ordering $\Tr(\l^{1}\l^{2}\l^{3}\l^{4}\l^{5})$. They are
 \beqa
A''_k(A_5T_1T_2T_3T_4)&=&(i\alpha'T_9)(k_3\cdot\xi_5-k_4\cdot\xi_5)\frac{1}{s_{34}}
\left(\frac{}{}\Tr(\l^{1}\l^{2}\l^{3}\l^{4}\l^{5})+\cdots\right)\nonumber\\
A''_k(A_5T_1T_4T_2T_3)&=&-(i\alpha'T_p)(k_2\cdot\xi_5-k_3\cdot\xi_5)\frac{1}{s_{23}}
\left(\Tr(\l^{1}\l^{2}\l^{3}\l^{4}\l^{5})+\cdots\right)\nonumber
\eeqa The second amplitude does not have $k_1\cdot\xi_5$
contribution, however, using conservation of momentum, the term in
the first amplitude above which is proportional to $k_4\cdot\xi_5$
has $k_1\cdot\xi_5$ contribution.  The total terms in fig.1 which
has coefficients $\Tr(\l^{1}\l^{2}\l^{3}\l^{4}\l^{5})$ and
$k_1\cdot\xi_5$ are then \beqa {\cal
A}_k^{TTTTA}&=&(i\alpha'T_9)k_1\cdot\xi_5\left(\frac{4}{s_{45}+\frac{1}{2}}+\frac{4}{s_{15}+\frac{1}{2}}-\frac{4}{s_{23}}
+\frac{4s_{23}}{(s_{45}+\frac{1}{2})s_{12}}+\frac{4s_{23}}{(s_{15}+\frac{1}{2})s_{34}}\right.\nonumber\\&&+
\frac{4s_{34}}{(s_{15}+\frac{1}{2})s_{23}}+\frac{4s_{12}}{(s_{45}+\frac{1}{2})s_{23}}+
\frac{4s_{23}}{s_{34}s_{12}}+\frac{4}{(s_{45}+\frac{1}{2})s_{12}}+\frac{4}{(s_{15}+\frac{1}{2})s_{34}}
\nonumber\\&&\left.+\frac{4}{(s_{15}+\frac{1}{2})s_{23}}+\frac{4}{(s_{45}+\frac{1}{2})s_{23}}+
\frac{4}{s_{34}s_{12}}\right)\Tr(\l^{1}\l^{2}\l^{3}\l^{4}\l^{5})+\cdots\labell{Ak}\eeqa
This is exactly the leading order terms of the non-abelian
expansion of the S-matrix element in the string theory side
\reef{Atttta}. This confirms that our prescription for expanding
the string theory S-matrix element produces correctly the
non-abelian expansion, an expansion whose first leading order
terms are reproduced by the non-abelian kinetic terms. The next
leading order terms in \reef{Atttta} are proportional to $\z(2)$.
We now turn to the Feynman amplitudes in field theory that are
proportional to $\z(2)$.

\subsection{$\zeta(2)$ order terms}

There are two main Feynman diagrams that are produced by one
vertex from the terms in the second or third line of
\reef{expandL} and one vertex from the kinetic terms. They appear
in fig.4. We write the amplitude as \beqa {\cal
A}^{TTTTA}_{\z(2)}&=&A_{\z(2)}+A'_{\z(2)}\nonumber\eeqa where
$A_{\z(2)}$ corresponds to diagram (a) and $A'_{\z(2)}$
corresponds to diagram (b). The Feynman amplitude of diagram (a)
is given by \beqa A_{\z(2)}(T_1T_2T_3T_4A_5)&=&{
V}_{\alpha}(T_1T_2T_3T)G_{\alpha\beta}(T){
V}_{\beta}(TT_4A_5)\labell{amp4}\eeqa The vertex of four tachyons
should be read from the different terms in the second line of
\reef{expandL}. To do this, one should first perform the symmetric
trace, \ie
\begin{figure}
{ \epsfbox {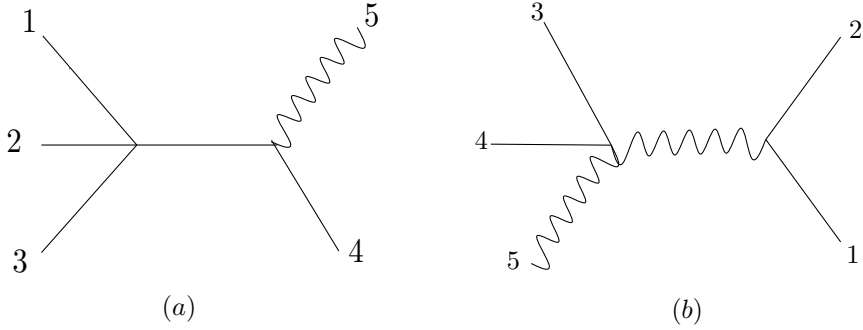}}
 \caption{The Feynman diagrams  at the $\zeta(2)$ order.}
\end{figure}
\beqa {\cal
L}^{TTTT}&=&-T_9(2\pi\alpha')^2\Tr\left(\frac{m^4}{8}T^4+\frac{m^2}{4}\left(\frac{2}{3}TT\prt_aT\prt^aT
+\frac{1}{3}T\prt_aTT\prt^aT\right)\right.\nonumber\\&&
-\left.\frac{1}{24}\left(2\prt_aT\prt^aT\prt_bT\prt^bT+\prt_aT\prt_bT\prt^aT\prt^bT\right)\right)\labell{Ltttt}
\eeqa  Now we can read the vertex of three on-shell tachyons and
one off-shell tachyon from the above couplings. In this section,
we choose to write the vertices in term of trace of $\l$'s. So the
vertex of four tachyons  contains terms that have group factors
$\Tr(\l^1\l^2\l^3\L^{\alpha})$, $\Tr(\l^2\l^1\l^3\L^{\alpha})$,
$\Tr(\l^2\l^3\l^1\L^{\alpha})$, $\Tr(\l^3\l^2\l^1\L^{\alpha})$,
$\Tr(\l^1\l^3\l^2\L^{\alpha})$, and
$\Tr(\l^3\l^1\l^2\L^{\alpha})$. However, after replacing them in
\reef{amp4}, only the first factor will produce the desired
ordering $\Tr(\l^1\l^2\l^3\l^4\l^5)$. So, we  consider only the
terms in the vertex that have factor
$\Tr(\l^1\l^2\l^3\L^{\alpha})$, \ie \beqa {
V}_{\alpha}(T_1T_2T_3T)&=&T_9i(2\pi\alpha')^2
\Tr(\lambda^{1}\lambda^{2}\lambda^{3}\Lambda^{\alpha})\nonumber\\&&
\times\left(\frac{m^4}{2}-\frac{m^2}{6} (-p^2+k_1\cdot
k_2+k_2\cdot k_3+k_1\cdot k_3)\right.\nonumber\\&&
\left.-\frac{1}{6}(k_1\cdot k_2 k_3\cdot p+k_2\cdot k_3 k_1\cdot
p+k_1\cdot k_3 k_2\cdot p)\right)+\cdots\nonumber\eeqa where $p$
is momentum of off-shell tachyon (see fig.5).  Replacing
\begin{figure}
{ \epsfbox {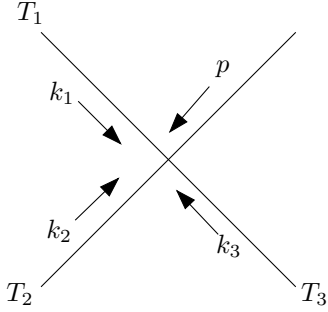}}
 \caption{ Vertex of three on-shell tachyons and one off-shell tachyon.  }
\end{figure}
this and the other vertex and propagator in \reef{amp4}, one finds
\beqa A_{\z(2)}(T_1T_2T_3T_4A_5)&=&-i \alpha'(8T_9\pi^2)k_4\cdot
\xi_5^{\alpha_5}(\frac{1}{s_{45}+\frac{1}{2}})
\Tr(\lambda^{1}\lambda^{2}\lambda^{3}\Lambda^{\alpha})if^{\alpha\alpha_4\alpha_5}\z_4^{\alpha_4}\nonumber\\&&
\times\left(\alpha'^2\frac{m^4}{2}-\alpha'^2\frac{m^2}{6}
(-p^2+k_1\cdot k_2+k_2\cdot k_3+k_1\cdot k_3)\right.\nonumber\\&&
-\frac{1}{6}\left[k_1\cdot k_2 (k_3\cdot
p+k_3\cdot k_5)+k_2\cdot k_3 (k_1\cdot p+k_1\cdot k_5)\right.\nonumber\\
&&\left.\left.\quad\quad+k_1\cdot k_3 (k_2\cdot p+k_2\cdot
k_5)\right] \frac{}{}\right)+\cdots\nonumber\eeqa Keeping the
ordering $\Tr(\l^1\l^2\l^3\l^4\l^5)$ and factor $k_1\cdot\xi_5$,
one finds \beqa
A_{\z(2)}(T_1T_2T_3T_4A_5)&=&i\alpha'T_9\z(2)\Tr(\lambda^{1}\lambda^{2}\lambda^{3}\lambda^{4}\lambda^{5})
 k_1\cdot \xi_5\nonumber\\&&\times
 \left(-\frac{4s_{12}^2}{s_{45}+\frac{1}{2}}-\frac{4s_{12}s_{23}}{s_{45}+\frac{1}{2}}
 -\frac{4s_{23}^2}{s_{45}+\frac{1}{2}}-\frac{8s_{12}}{s_{45}+\frac{1}{2}}
 -\frac{8s_{23}}{s_{45}+\frac{1}{2}}-\frac{4}{s_{45}+\frac{1}{2}}\right.\nonumber\\&&\left.
 +4s_{12}+4s_{23}+8\frac{}{}\right)+\cdots\labell{amp6}\eeqa
where  we have  used relation \reef{cons1} and $\pi^2/6=\z(2)$.

 The other distinct  diagram that produces the
ordering
 $\Tr(\lambda^{1}\lambda^{2}\lambda^{3}\lambda^{4}\lambda^{5})$
 is
   \beqa A_{\z(2)}(T_2T_3T_4T_1A_5)&=&-i \alpha'
 T_9\z(2)\Tr(\lambda^{1}\lambda^{2}\lambda^{3}\lambda^{4}\lambda^{5})
 (-k_1\cdot \xi_5)\nonumber\\&&
 (-\frac{4s_{23}^2}{s_{15}+\frac{1}{2}}-\frac{4s_{34}s_{23}}{s_{15}+\frac{1}{2}}
 -\frac{4s_{34}^2}{s_{15}+\frac{1}{2}}-\frac{8s_{23}}{s_{15}+\frac{1}{2}}
 -\frac{8s_{34}}{s_{15}+\frac{1}{2}}-\frac{4}{s_{15}+\frac{1}{2}}\nonumber\\&&
 +4s_{23}+4s_{34}+8)+\cdots
\labell{amp7}\eeqa where again dots refer to the terms that have
coefficient $k_2\cdot\xi_5$, $k_3\cdot\xi_5$, and have group
factor other than
$\Tr(\lambda^{1}\lambda^{2}\lambda^{3}\lambda^{4}\lambda^{5})$.

We now consider the diagram (b) in fig. 4. The amplitude is given
by \beqa A'_{\z(2)}(A_5T_1T_2T_3T_4)&=&{
V}^a_{\alpha}(A_5T_1T_2A)G^{ab}_{\alpha\beta}(A)
V^b_{\beta}(AT_3T_4)\labell{amp5}\eeqa where the vertex of two
tachyons and two gauge fields should be read from the terms in the
third line of \reef{expandL}. Performing the  symmetric trace,
they can be written as \beqa {\cal
L}^{AATT}&\!\!\!\!=\!\!\!\!&-T_9\frac{(2\pi\alpha')^3}{2}\Tr\left((\frac{1}{3}(\prt^aT\prt_bTF^{bc}F_{ca})
+\frac{1}{3}(\prt^aT\prt_bTF^{ca}F_{bc})\right.\nonumber\\&&+\frac{1}{6}(\prt^aTF^{bc}\prt_bTF_{ca})
+\frac{1}{6}(\prt^aTF^{ca}\prt_bTF_{bc})\nonumber\\&&-\frac{1}{4}(\frac{2}{3}(\prt^aT\prt_aTF^{bc}F_{cb})
+\frac{1}{3}(\prt^aTF^{bc}\prt_aTF_{cb}))\nonumber\\&&\left.
-\frac{m^2}{4}(\frac{2}{3}(TTF^{ab}F_{ba})+\frac{1}{3}(TF^{ab}TF_{ba}))\right)\labell{Laatt}\\&\!\!\!\!\equiv\!\!\!\!&{\cal
L}^{AATT}_1+{\cal L}^{AATT}_2+{\cal L}^{AATT}_3+{\cal
L}^{AATT}_4+{\cal L}^{AATT}_5+{\cal L}^{AATT}_6+{\cal
L}^{AATT}_7+{\cal L}^{AATT}_8\nonumber\eeqa The possible terms in
the above interaction that produce in the amplitude \reef{amp5}
the ordering
$\Tr(\lambda^{1}\lambda^{2}\lambda^{3}\lambda^{4}\lambda^{5})$ are
${\cal L}^{AATT}_1+{\cal L}^{AATT}_2+{\cal L}^{AATT}_5+{\cal
L}^{AATT}_7$. One can find the vertex of two tachyons and two
gauge fields from these terms. It contains terms that have group
factors $\Tr(\l^5\l^1\l^2\L^{\alpha})$,
$\Tr(\l^1\l^2\l^5\L^{\alpha})$,  $\Tr(\l^2\l^1\l^5\L^{\alpha})$,
and $\Tr(\l^5\l^2\l^1\L^{\alpha})$. However, after replacing them
in \reef{amp5}, only the first factor will produce the desired
ordering $\Tr(\l^1\l^2\l^3\l^4\l^5)$. So, we  consider only the
terms in the vertex that have factor
$\Tr(\l^5\l^1\l^2\L^{\alpha})$. By replacing it and the other
vertex from kinetic term in \reef{amp5}, one finds that the
amplitude has terms proportional to $k_1\cdot\xi_5$,
$k_2\cdot\xi_5$, $k_3\cdot\xi_5$, and $k_4\cdot\xi_5$. The last
factor should be written as $-(k_1+k_2+k_3)\cdot\xi_5$.
To compare the result with the terms in the string theory side
\reef{Atttta}, we  consider only terms that have coefficient
$k_1\cdot\xi_5$ and group factor
$\Tr(\lambda^{1}\lambda^{2}\lambda^{3}\lambda^{4}\lambda^{5})$.
The result is
 \beqa A'_{\z(2)}(A_5T_1T_2T_3T_4)&=&i\alpha'T_9\z(2) k_1\cdot
\xi_5\Tr(\lambda^{1}\lambda^{2}\lambda^{3}\lambda^{4}\lambda^{5})
\nonumber\\&&\times\left(-\frac{4s_{23}(s_{15}+\frac{1}{2})}{s_{34}}-\frac{4s_{12}s_{23}}{s_{34}}
-\frac{4(s_{15}+\frac{1}{2})}{s_{34}}-\frac{4s_{12}}{s_{34}}\right.
\nonumber\\&&\left.
-2(s_{15}+\frac{1}{2})-2s_{12}+2s_{34}+4s_{23}+4\right)+\cdots\labell{amp8}\eeqa
Another distinct diagram that produces the ordering
$\Tr(\lambda^{1}\lambda^{2}\lambda^{3}\lambda^{4}\lambda^{5})$ and
has coefficient  $k_1\cdot\xi_5$ is
 \beqa A'_{\z(2)}(T_3T_4A_5T_1T_2)&=&{
V}^a_{\alpha}(T_3T_4A_5A)G^{ab}_{\alpha\beta}(A)
V^b_{\beta}(AT_1T_2)\nonumber\\
&=&i\alpha'T_9\z(2) k_1\cdot
\xi_5\Tr(\lambda^{1}\lambda^{2}\lambda^{3}\lambda^{4}\lambda^{5})
\nonumber\\&&\times\left(-\frac{4s_{23}(s_{45}+\frac{1}{2})}{s_{12}}-\frac{4s_{23}s_{34}}{s_{12}}
-\frac{4(s_{45}+\frac{1}{2})}{s_{12}}
-\frac{4s_{34}}{s_{12}}\right.\nonumber\\&&\left.
-2(s_{45}+\frac{1}{2})+2s_{12}+4s_{23}-2s_{34}+4\right)+\cdots\labell{amp9}\eeqa
In above amplitude again only ${\cal L}^{AATT}_1+{\cal
L}^{AATT}_2+{\cal L}^{AATT}_5+{\cal L}^{AATT}_7$ has contribution.
The remaining terms in \reef{Laatt}, \ie ${\cal L}^{AATT}_3+{\cal
L}^{AATT}_4+{\cal L}^{AATT}_6+{\cal L}^{AATT}_8$ has contribution
into the following amplitude: \beqa
 A'_{\z(2)}(T_4A_5T_1T_2T_3)&=&{
V}^a_{\alpha}(T_4A_5T_1A)G^{ab}_{\alpha\beta}(A){
V}^b_{\beta}(AT_2T_3)\nonumber\\
&=&i\alpha'T_9\z(2)k_1\cdot
\xi_5Tr(\lambda^{1}\lambda^{2}\lambda^{3}\lambda^{4}\lambda^{5})\nonumber
\\&&\times\left(\frac{4(s_{15}+\frac{1}{2})(s_{45}+\frac{1}{2})}{s_{23}}-\frac{4(s_{15}+\frac{1}{2})s_{12}}{s_{23}}
-\frac{4(s_{45}+\frac{1}{2})s_{34}}{s_{23}}\right.\labell{amp10}\\&&\left.-\frac{4(s_{15}+\frac{1}{2})}{s_{23}}
-\frac{4(s_{45}+\frac{1}{2})}{s_{23}}-2(s_{15}+\frac{1}{2})-2(s_{45}+\frac{1}{2})\right)+\cdots\nonumber\eeqa
Comparing terms in ${\cal A}_{\z(2)}^{TTTTA}$ which have
coefficients
$\Tr(\lambda^{1}\lambda^{2}\lambda^{3}\lambda^{4}\lambda^{5})$ and
$k_1\cdot\xi_5$, \ie equations \reef{amp6}, \reef{amp7},
\reef{amp8}, \reef{amp9}, and \reef{amp10}, with the non-abelian
expansion of string theory S-matrix element \reef{Atttta}, one
finds that the poles of \reef{Atttta} which are proportional to
$\z(2)$ are exactly reproduced by the above field theory
amplitudes. This confirms that while the leading order terms of
the non-abelian expansion of S-matrix element \reef{Atttta} are
reproduced by non-abelian kinetic terms \reef{Ak}, the next
leading order terms of the  S-matrix element are reproduced by
non-abelian tachyon DBI action. In particular,  it confirms the
four-tachyon couplings in \reef{Ltttt}. These terms appear in one
of the vertex in diagram (a) in fig.4. Since one of the legs of
the vertex is off-shell, there is no on-shell ambiguity in these
terms. There is such ambiguity in finding them from studying the
non-abelian expansion of the S-matrix element of four tachyons
\cite{garousi}. In that calculation, the couplings \reef{Ltttt}
appear as contact term in field theory, hence, there is on-shell
ambiguity\footnote{See \cite{VF}, for an off-shell $T^4$ coupling
in cubic string field theory.}.

We have seen that the tachyon and massless poles in the
non-abelian expansion of the string theory S-matrix element
\reef{Atttta} are exactly reproduced by the tachyon and massless
poles of the non-abelian DBI action \reef{qt1}. One can even show
that the contact terms of \reef{Atttta} which appear in the next
leading order terms, \ie the contact terms which are proportional
to $\z(2)$, can be reproduced by non-abelian DBI action. To this
end, we subtract the field theory amplitudes ${\cal
A}_k^{TTTTA}+{\cal A}_{\z(2)}^{TTTTA}$ from  \reef{Atttta}, \ie
\beqa {\cal A}^{TTTTA}-{\cal A}_k^{TTTTA}-{\cal
A}_{\z(2)}^{TTTTA}&=&i\alpha'T_9\z(2) k_1\cdot
\xi_5\Tr(\lambda^{1}\lambda^{2}\lambda^{3}\lambda^{4}\lambda^{5})\left(\frac{}{}4s_{23}+8\frac{}{}
\right)+\cdots \nonumber\eeqa where dots refer to the terms with
coefficients $\z(3)$, $\z(4)$, and so on. One can easily check
that the above contact terms are exactly  the four tachyons and
one gauge field couplings in the second line of \reef{expandL}.
This ends our illustration of consistency between  the non-abelian
expansion of  S-matrix element \reef{Atttta}  and the non-abelian
DBI action \reef{qt1}.

The next order terms in the expansion \reef{Atttta} are
proportional to $\z(3)$. They contains massless, tachyon poles,
and contact terms. Like the terms in the kinetic order and $\z(2)$
order, the different terms of $\z(3)$-order are not of the same
$\alpha'$ order. We speculate that the $\z(3)$-order terms in
\reef{Atttta} are  reproduced by second derivative of $T$, \ie
couplings that include $\prt\prt T$. Similarly, in studying the
non-abelian expansion of S-matrix element of four tachyons
\cite{garousi}, the kinetic order terms and $\z(2)$-order terms
are shown to be reproduced by tachyon DBI action. In that case
also the next leading order terms which are only contact terms are
proportional to $\z(3)$. Those terms may also be written as second
derivative of $T$, \eg $\alpha'^2T^2\prt_a\prt_b T\prt^a\prt^b T$
or $\alpha'^3\prt_a T\prt^a T\prt_b\prt_c T\prt^b\prt ^cT$.
However, there are on-shell ambiguity in these couplings, \eg
$\alpha'^2T^2\prt_a\prt_b T\prt^a\prt^b
T\sim\alpha'^3T\prt_c\prt^cT\prt_a\prt_b T\prt^a\prt^b T$. This
ambiguity can be fixed by studying them in the S-matrix element of
four tachyons and one gauge field in which the $\z(3)$-order terms
have tachyon poles. It would be interesting to explicitly perform
these calculations to find the second-derivative corrections to
the tachyon DBI action. Another interesting calculation is to
evaluate the S-matrix element of six tachyons. Comparing the
result with the S-matrix element of six massless scalars
\cite{DO}, one would find the non-abelian expansion of the tachyon
amplitude. This calculation would fix the $T^6$ term in the
tachyon potential \reef{potential}.

{\bf Acknowledgement}: We would like to thank R. Medina, T.
Brandt, D. Maitre, M. Yu. Kalmykov and R. Zare for valuable
conversations.
\newpage
\appendix

\section{appendix }
The momentum dependent factors $ a',b',c'$ in  the tachyon
scattering amplitude \reef{amp}, and $a_1,a_2,a_3$, $b_1,b_2,b_3$,
$c_1,c_2,c_3$ in the scalar scattering amplitude \reef{amp01} are
the double integral that left over after fixing position of three
vertex operators in \reef{amp0} and \reef{amp00}, respectively. In
general, using Mathematica software, one can write the following
double integral in terms of two Beta functions and one
hypergeometric function: \beqa \int_0^1 d x_3\int_0^{x_3} d x_2
x_3^{-s_{13}-n_{13}} (1-x_3)^{ -s_{34}-n_{34}}
{x_2}^{-s_{12}-n_{12}}(1-x_2)^{-s_{24}-n_{24}}
 (x_3-x_2)^{-s_{23}-n_{23}}\nonumber\\=\beta(1-s_{12}-n_{12},1-s_{23}-n_{23})
 \beta(2-s_{12}-s_{13}-s_{23}-n_{12}-n_{13}-n_{23},1-s_{34}-n_{34})\nonumber\\\times
{}_{3}F_2(s_{24}+n_{24},1-s_{12}-n_{12},2-s_{12}-s_{13}-s_{23}-n_{12}-n_{13}-n_{23};\nonumber\\
2-s_{12}-s_{23}-n_{12}-n_{23},3-s_{34}-s_{12}-s_{13}-s_{23}-n_{34}-n_{12}-n_{13}-n_{23};1)\nonumber\eeqa
The integers  $n_{ij }$ for each factor are listed in the
following table:

 {\setlength{\tabcolsep}{4pt}
 {\begin{tabular} {|c|c|c|c|c|c|} \hline momentum factors &
$n_{24}$ & $n_{23}$ & $n_{12}$ & $n_{34}$ &  $n_{13}$
\\ \hline
$a_1$ & $0$ & $1$ & $1$ & $1$ & $0$
\\
 $a_2$ & $1$ & $1$ & $0$ & $0$ & $1$
\\
$a_3$ & $0$ & $2$ & $0$ & $0$ & $0$
\\
$b_1$ & $0$ & $0$ & $1$ & $1$ & $1$
\\
$b_2$ & $1$ & $0$ & $0$ & $0$ & $2$
\\
$b_3$ & $0$ & $1$ & $0$ & $0$ & $1$
\\
$c_1$ & $0$ & $0$ & $2$ & $1$ & $0$
\\
$c_2$ & $1$ & $0$ & $1$ & $0$ & $1$
\\
$c_3$ & $0$ & $1$ & $1$ & $0$ & $1$
\\
$a'$ & $1$ & $1$ & $2$ & $1$ & $1$
\\
$b'$ & $1$ & $2$ & $1$ & $1$ & $1$ \\
$c'$ & $2$ & $1$ & $1$ & $1$ & $1$
 \\ \hline
\end{tabular}}\\\\\\
The factors $a_1,~a_2,~a_3, ~b_1,~b_2,~b_3,~c_1,~c_2,~c_3$ are the
same factors that appear in the S-matrix element of five gauge
fields \cite{medina}. There, they have been
named $L_2$, $L_7$, $~K_6$, $~L_3$, $~L_6$, $~K_5$, $~L_5$, $~L_1$, $~K_4$, respectively .\\

The factor in the tachyon amplitude \reef{amp} which is multiplied
by the kinematic factor $k_1\cdot\xi_5$ in which we are interested
is \beqa &&a'=\beta(-s_{12},-s_{23}-1)
\beta(-s_{45}-\frac{1}{2},-s_{34})\nonumber\\&&\times
{}_{3}F_2(-s_{12},-s_{45}-\frac{1}{2},s_{15}-s_{23}-s_{34}-\frac{1}{2};
-s_{12}-s_{23}-1,-s_{34}-s_{45}-\frac{1}{2};1)\nonumber\eeqa
We are going to find the non-abelian expansion of this factor.  
Consider the following function:

 \beqa a=\beta(l_1,l_2)\beta(l_3,l_4){}_3F_2(m_1,m_2,m_3;n_1,n_2;1)\nonumber\eeqa
If one chooses appropriate value for $l_i$, $m_i$, and $n_i$, the
above function converts to $a'$ or to one of the other momentum
factors. For the non-abelian expansion of $a'$, one should expand
the above function around three different points \reef{expanda}.
Expansion of Beta function is
 straightforward, using Mathematica,
 \beqa
   \lim_{l_1, l_2, l_3, l_4\rightarrow 0} \beta(l_1,l_2)\beta(l_3,l_4)&\!\!\!=\!\!\!&
\frac{(l_1+l_2)(l_3+l_4)}{l_1l_2l_3l_4}-\zeta(2)
\frac{(l_1+l_2)(l_3+l_4)(l_1l_2+l_3l_4)}{l_1l_2l_3l_4}+\cdots\nonumber\\
  \lim_{l_1, l_4\rightarrow1, l_2, l_3\rightarrow 0} \beta(l_1,l_2)\beta(l_3,l_4)&\!\!\!=\!\!\!&
\frac{(1+l_1+l_2)(1+l_3+l_4)}{(1+l_1)l_2l_3(1+l_4)}\nonumber\\&&-\zeta(2)
\frac{(1+l_1+l_2)(1+l_3+l_4)(l_2+l_3+l_1l_2+l_3l_4)}{(1+l_1)l_2l_3(1+l_4)}+\cdots\nonumber\\
  \lim_{l_1, l_4\rightarrow1, l_2\rightarrow -1, l_3\rightarrow 0} \beta(l_1,l_2)\beta(l_3,l_4)&\!\!\!=\!\!\!&
\frac{(l_1+l_2)(1+l_3+l_4)}{(1+l_1)(-1+l_2)l_3(1+l_4)}\nonumber\\&&-\zeta(2)
\frac{(l_1+l_2)(1+l_3+l_4)(-1+l_2-l_1+l_3+l_1l_2+l_3l_4)}{(1+l_1)(-1+l_2)l_3(1+l_4)}\nonumber\\
&&+\cdots\labell{beta}
\eeqa where dots refer to the terms that have coefficients
$\z(3)$, $\z(4)$ and so on. The nontrivial part of expansion is
the  expansion of Hypergeometric function. This function  appears
also in higher terms of the $\varepsilon$-expansion of massive
Feynman diagrams \cite{AI}.  There are different approaches for
expanding this function \cite{YK,MY}. In these references, one can
find many useful techniques for expansion explicitly the
Hypergeometric function.  Fortunately, there is a package for
expanding  the Hypergeometric functions \cite{TH}. Using this
package, one finds the following expansion for Hypergeometric
function around the three point \reef{expanda}:

\beqa
  &&\lim_{m_1, m_2, m_3, n_1, n_2\rightarrow 0}{}_3F_2(m_1,m_2,m_3;n_1,n_2;1)\!=\!
  \frac{n_1n_2(-m_2-m_3+n_1+n_2)+m_1(m_2m_3-n_1n_2)}
{n_1n_2(-m_1-m_2-m_3+n_1+n_2)}\nonumber\\&& -\z(2)\frac{m_1m_2m_3}
{n_1n_2(m_1+m_2+m_3-n_1-n_2)}
(n_1^2-m_3n_1+n_2n_1+n_2^2\nonumber\\&&+m_2(m_3-n_1-n_2)+m_1(m_2+m_3-n_1-n_2)-m_3n_2)+\cdots\nonumber\\
  &&\lim_{m_1, m_3, n_1, n_2\rightarrow 1,m_2\rightarrow 0}{}_3F_2(m_1,m_2,m_3;n_1,n_2;1)=
\frac{1}{(n_1+1)(n_2+1)(n_1+n_2-m_1-m_2-m_3)}\nonumber\\&&
\times(n_1+n_2+(n_1+n_2-m_2)(n_1+n_2+n_1n_2)+
m_3(m_2-(n_1+1)(n_2+1)\nonumber\\&&+m_1(m_2(m_3+1)-(n_1+1)(n_2+1)))\nonumber\\&&+\zeta(2)
\frac{m_2(1+m_1)(1+m_3)}{(n_1+1)(n_2+1)(n_1+n_2-m_1-m_2-m_3)}
(n_2^2+n_2(1+n_1-m_2-m_3)\nonumber\\&&+(m_2-n_1-1)(m_3-n_1)+m_1(m_2+m_3-n_1-n_2-1))+\cdots\nonumber\\
 &&\lim_{m_1, n_2\rightarrow 1, m_2,m_3,n_1\rightarrow0}{}_3F_2(m_1,m_2,m_3;n_1,n_2;1)=
\frac{1}{n_1(1+n_2)(n_1+n_2-m_1-m_2-m_3)}\nonumber\\&&\times
(n_1(1+n_2)(n_1+n_2-m_1-m_3)+m_2(m_3(1+m_1)-n_1(1+n_2)))\nonumber\\&&
+\zeta(2)\frac{m_2m_3(1+m_1)}{n_1(1+n_2)(m_1+m_2+m_3-n_1-n_2)}(n_1^2+n_2^2+n_1n_2-n_1m_3\nonumber\\&&
+m_2(m_3-n_1-n_2)+m_1(m_2+m_3-n_1-n_2-1)-m_3n_2+n_2)+\cdots\labell{F32}
 \eeqa
 Using the expansions \reef{beta} and \reef{F32}, and choosing the
 specific value of $l_i$, $m_i$ and $n_i$ corresponding to the
 momentum factor $a'$, one finds the following non-abelian
 expansion:
\beqa a'_1&=&\lim{}_{s_{23}\rightarrow-1,~s_{45},s_{15}\rightarrow
-\frac{1}{2},~s_{12},s_{34}\rightarrow
0}~~a'=\frac{1}{s_{12}(s_{45}+\frac{1}{2})}+\frac{1}{s_{34}(s_{15}+\frac{1}{2})}\nonumber\\&&
+\frac{1}{s_{12}s_{34}}
+\frac{1}{(s_{23}+1)(s_{45}+\frac{1}{2})}+\frac{1}{(s_{23}+1)(s_{15}+\frac{1}{2})}\nonumber
\\&& -\zeta(2)\left(\frac{(s_{15}+\frac{1}{2})}{(s_{23}+1)}+\frac{(s_{15}+\frac{1}{2})}{s_{34}}
+\frac{(s_{45}+\frac{1}{2})}{(s_{23}+1)}+\frac{s_{12}}{(s_{45}+\frac{1}{2})}\right.\nonumber
\\&&\left.+
\frac{s_{34}}{s_{12}}+\frac{(s_{23}+1)}{(s_{45}+\frac{1}{2})}+\frac{(s_{45}+\frac{1}{2})}{s_{12}}
+\frac{s_{12}}{s_{34}}+\frac{(s_{23}+1)}{(s_{15}+\frac{1}{2})}+\frac{s_{34}}{(s_{15}+\frac{1}{2})}
-3\right)+\cdots\nonumber\\
a'_2&=&\lim{}_{s_{12},s_{23},s_{34}\rightarrow-1,~s_{45},s_{15}\rightarrow
-\frac{1}{2}}~~a'=
\frac{1}{(s_{15}+\frac{1}{2})(s_{23}+1)}+\frac{1}{(s_{45}+\frac{1}{2})(s_{23}+1)}
\nonumber\\&&-\zeta(2)\left(\frac{(s_{12}+1)}{(s_{45}+\frac{1}{2})}+\frac{(s_{34}+1)}{(s_{15}+\frac{1}{2})}+
\frac{(s_{45}+\frac{1}{2})}{(s_{23}+1)}+\frac{(s_{15}+\frac{1}{2})}{(s_{23}+1)}-1\right)+\cdots\nonumber\\
a'_3&=&\lim{}_{s_{12},s_{34}\rightarrow-1,~s_{45},s_{15}\rightarrow
-\frac{1}{2}~s_{23}\rightarrow
0}~~a'=-\frac{(s_{12}+1)}{(s_{45}+\frac{1}{2})}+\frac{(s_{12}+1)}{s_{23}(s_{45}+\frac{1}{2})}
-\frac{(s_{34}+1)}{(s_{15}+\frac{1}{2})}\nonumber\\&&-\frac{s_{23}}{(s_{15}+\frac{1}{2})}-
\frac{s_{23}}{(s_{45}+\frac{1}{2})}+\frac{1}{(s_{45}+\frac{1}{2})}+
\frac{1}{(s_{15}+\frac{1}{2})}+\frac{(s_{34}+1)}{s_{23}(s_{15}+\frac{1}{2})}-\frac{1}{s_{23}}+1\nonumber\\&&
+\frac{s_{23}(1+s_{12})}{(s_{45}+\frac{1}{2})}+\frac{s_{23}(s_{34}+1)}{(s_{15}+\frac{1}{2})}-s_{23}
+\frac{s_{23}^2}{(s_{15}+\frac{1}{2})}+\frac{s_{23}^2}{(s_{45}+\frac{1}{2})}+
\zeta(2)\left(-\frac{(1+s_{12})^2}{(s_{45}+\frac{1}{2})}\right.\nonumber\\
&&-\frac{s_{23}(1+s_{12})}{(s_{45}+\frac{1}{2})}-\frac{(s_{15}+\frac{1}{2})(s_{12}+1)}{s_{23}}
-\frac{s_{23}(s_{34}+1)}{(s_{15}+\frac{1}{2})}+\frac{(s_{15}+\frac{1}{2})(s_{45}+\frac{1}{2})}{s_{23}}
\nonumber\\&&\left.-\frac{(s_{34}+1)(s_{45}+\frac{1}{2})}{s_{23}}-\frac{(s_{34}+1)^2}{(s_{15}+\frac{1}{2})}
+s_{12}+s_{23}-s_{45}-s_{15}+s_{34}+1 \right)+\cdots\nonumber\eeqa
Inserting the above expansion in \reef{limit}, one finds the
non-abelian expansion of the tachyon amplitude \reef{Atttta}.



\begin{thebibliography}{99}
\bibitem
{mgas}{M. Gutperle and A. Strominger,  JHEP {\bf 0204},018 (2002)
[arXiv: hep-th/0202210].}
\bibitem
{asen1}{A. Sen,  JHEP {\bf 0204}, 048 (2002) [arXiv:
hep-th/0203211].}
\bibitem
{asen2}{A. Sen,   JHEP {\bf 0207}, 065 (2002) [arXiv:
hep-th/0203265].}
\bibitem
{asen3}{A. Sen,  Mod. Phys. Lett. A {\bf 17}, 1797 (2002) [arXiv:
hep-th/0204143].}
\bibitem
{asen4}{A. Sen,  JHEP {\bf 0210}, 003 (2002) [arXiv:
hep-th/0207105].}
\bibitem
{asen5}{A. Sen, Int. J. Mod. Phys. A {\bf 18} (2003) 4869 [ arXiv:
hep-th/0209122].}
\bibitem
{pmas}{P. Mukhopadhyay and A. sen,  JHEP {\bf 0211}, 047 (2002)
[arXiv: hep-th/0208142].}
\bibitem
{as}{A. Strominger, ``Open string creation by S-branes,'' arXiv:
hep-th/0209090.}
\bibitem
{flanst}{F. Larsen, A. Naqvi and S. Terashima,  JHEP {\bf 0302},
039 (2003) [arXiv: hep-th/0212248].}
\bibitem
{mgas1}{M. Gutperle and A. Strominger, Phys. Rev. D {\bf 67},
126002 (2003) [arXiv: hep-th/0301038].}
\bibitem
{amas}{A. Maloney, A. Strominger and X. Yin, JHEP {\bf 0310}
(2003) 048 [ arXiv: hep-th/0302146].}
\bibitem
{toss}{T. Okuda and S. Sugimoto,  Nucl. Phys. B {\bf 647}, 101
(2002) [arXiv: hep-th/0208196].}
\bibitem
{srss}{S. J. Rey and S. Sugimoto, Phys. Rev. D {\bf 67}, 086008
(2003) [arXiv: hep-th/0301049].}
\bibitem
{srss1}{S. J. Rey and S. Sugimoto, Phys. Rev. D {\bf 68}, 026003
(2003)  [arXiv: hep-th/0303133].}
\bibitem
{nrc}{N.R. Constable and F. Larsen, JHEP {\bf 0306}, 017 (2003)
[arXiv: hep-th/0305177].}
\bibitem
{bcml}{B. Chen, M. Li and F.L. Lin, JHEP {\bf 0211}, 050 (2002)
[arXiv: hep-th/0209222].}
\bibitem
{ss}{S. Sugimoto and S. Terashima, JHEP {\bf 0207}, 025 (2002)
[arXiv: hep-th/0205085].}
\bibitem
{jam}{J.A. Minahan, JHEP {\bf 0207}, 030 (2002) [arXiv:
hep-th/0205098].}
\bibitem
{nm}{N. Moeller and B. Zwiebach, JHEP {\bf 0210}, 034 (2002)
[arXiv: hep-th/0207107].}
\bibitem
{mf}{M. Fujita and H. Hata, JHEP {\bf 0305}, 043 (2003) [arXiv:
hep-th/0304163].}
\bibitem
{iya}{I.Y. Aref'eva, L.V. Joukovskaya and A.S. Koshelev, JHEP {\bf
0309} (2003) 012 [ arXiv: hep-th/0301137].}
\bibitem
{nlhl}{N. Lambert, H. Liu and J. Maldacena, ``Closed strings from
decaying D-branes,'' [arXiv: hep-th/0303139].}
\bibitem
{dgni}{D. Gaiotto, N. Itzhaki and L. Rastelli, Nucl. Phys. B {\bf
688} (2004) 70 [ arXiv: hep-th/0304192].}
\bibitem
{jm}{J. McGreevy and Verlinde, JHEP {\bf 0312} (2003) 054 [ arXiv:
hep-th/0304224].}
\bibitem
{jm2}{J. McGreevy, J. Teschner and H. Verlinde, JHEP {\bf 0401}
(2004) 039 [ arXiv: hep-th/0305194].}
\bibitem
{irk}{I.R. Klebanov, J. Maldacena and N. Seiberg, JHEP {\bf 0307}
(2003) 045 [ arXiv: hep-th/0305159].}
\bibitem
{sen1}{A. Sen, Phys. Rev. D {\bf 68} (2003) 106003 [ arXiv:
hep-th/0305011].}

\bibitem
{jlk}{J.L. Karczmarek, H. Liu, J. Maldacena and A. Strominger,
JHEP {\bf 0311} (2003) 042 [ arXiv: hep-th/0306132].}
\bibitem
{sen2}{A. Sen, Phys. Rev. Lett. {\bf 91} (2003) 181601 [ arXiv:
hep-th/0306137].}
\bibitem
{VB}{V. Balasubramanian, E. Keski-Vakkuri, P. Kraus and A. Naqui,
Commun. Math. Phys. {\bf 257} (2005) 363 [arXiv: hep-th/0404039].}
\bibitem
{JS}{J. Shelton, JHEP {\bf 0501} (2005) 037 [arXiv:
hep-th/0411040].}
\bibitem
{NJ}{N. Jokela, E. Keski-Vakkuri and J. Majumder, Phys. Rev. D
{\bf 73} (2006) 046007 [arXiv: hep-th/0510205].}

\bibitem
{mg}{M. R. Garousi, Nucl. Phys. B {\bf 584}, 284 (2000)
[arXiv:hep-th/0003122].}
\bibitem
{ebmr}{E. A. Bergshoeff, M. de Roo, T. C. de Wit, E. Eyras and S.
Panda, JHEP {\bf 0005}, 009 (2000) [arXiv:hep-th/0003221];\\
J. Kluson, Phys. Rev. D {\bf 62}, 126003 (2000)
[arXiv:hep-th/0004106].}
\bibitem
{ASen}{A. Sen, Int. J. Mod. Phys. A {\bf 20}, 5513 (2005) [arXiv:
hep-th/0410103].}

\bibitem
{AC}{A. Connes, M.R. Douglas and A. Schwarz, JHEP {\bf 02} (1998)
003; [hep-th/9711162].}
\bibitem
{mrd}{M.R. Douglas, C. Hull, JHEP {\bf 02} (1998) 008;
[hep-th/9711165].}
\bibitem
{faha}{F. Ardalan, H. Arfaei and M.M. Sheikh-Jabbari, JHEP {\bf
02} (1999) 014; [hep-th/9810072].}

\bibitem
{csc}{C.S. Chu and P.M. Ho, Nucl. Phys. B {\bf 550} (1999) 151;
[hep-th/9812219].}
\bibitem
{NS}{N. Seiberg and E. Witten, JHEP {\bf  09 } (1999) 032 [
hep-th/9908142].}
\bibitem
{AAT}{A.A. Tseytlin, Nucl. Phys. B {\bf 276} (1987) 391.}
\bibitem
{AAT1}{A.A. Tseytlin, Phys. Lett. B {\bf 176} (1986) 92; R.R.
Metsaev, A.A. Tseytlin, Phys. Lett. B {\bf 185} (1987) 52.}

\bibitem
{ew1}{E. Witten, Nucl. Phys. B {\bf 460}, 335 (1996) [arXiv:
hep-th/9510135].}
\bibitem
{mohammad}{M.R. Garousi, JHEP {\bf 0501} (2005), 026 [arXiv:
hep-th/0411222]; K. Bitaghsir and M.R. Garousi, JHEP {\bf 0604}
(2006), 005 [arXiv: hep-th/0506055].}

\bibitem
{garousi}{M.R. Garousi, JHEP {\bf 0312} (2003) 036 [arXiv:
hep-th/0307197].}
\bibitem
{garousi1}{M.R. Garousi, Nucl. Phys. B {\bf 647}, 117 (2002)
[arXiv: hep-th/0209068].}
\bibitem
{garousi11}{M.R. Garousi,  JHEP {\bf 0304} (2003) 027 [arXiv:
hep-th/0303239].}
\bibitem
{garousi2}{M.R. Garousi,  JHEP {\bf 0305} (2003) 058 [arXiv:
hep-th/0304145].}
\bibitem
{DKV}{D. Kutasov, V. Niarchose, Nucl. Phys. B {\bf 666}, 56 (2003)
[arXiv: hep-th/0304045]; V. Niarchose, Phys. Rev. D {\bf 69},
106009 (2004) [arXiv: hep-th/0401066].}
\bibitem
{jp}{J. Polchinski, ``String theory,'' Cambridge University Press,
1998.}
\bibitem
{YK}{Y. Kitazawa, Nucl. Phys. B {\bf 289}, 599 (1987).}
\bibitem
{ZK}{Z. Koba and H.B. Nielsen, Nucl. Phys. B {\bf 10}, 633 (1969);
Nucl. Phys. B {\bf 12}, 517 (1969).}

\bibitem
{myers}{R.C. Myers, JHEP {\bf 9912} (1999) 022 [arXiv:
hep-th/9910053]; M.R. Garousi and R.C. Myers, JHEP {\bf 0011}
(2000) 032 [arXiv: hep-th/0010122].}
\bibitem
{DK}{D. Kutasov, M. Marino and G. Moore, ``Remarks on Tachyon
Condensation in Superstring Feild Theory,'' hep-th/0010108.}

\bibitem
{JAM}{J.A. Minahan and B. Zwiebach, JHEP {\bf 0103} (2001) 038
[arXiv: hep-th/0009246].}
\bibitem
{medina}{R. Medina, F.T. Brandt and F.R. Machado, JHEP {\bf 07}
(2002) 071 [arXiv:hep-th/0208121]; L.A. Barreiro and R. Medina,
JHEP {\bf 0311} (2003) 003 [arXiv: hep-th/0503182].}
\bibitem
{AI}{A.I. Davydychev and M. Yu. Kalmykov, Nucl. Phys. B {\bf 699},
3 (2004) [arXiv: hep-th/0303162].}

\bibitem
{MY}{M. Yu. Kalmykov, Nucl. Phys. Proc. Suppl. {\bf 135}, 280
(2004) [arXiv: hep-th/0406269]}.
\bibitem
{TH}{T. Huber and D. Maitre, Comput. Phys. Commun. {\bf 175}, 122
(2006) [arXiv: hep-ph/0507094].}
\bibitem
{VF}{V. Forini, G. Grignani and G. Nardelli, JHEP {\bf 0604}
(2006) 053 [arXiv: hep-th/0603206].}

\bibitem
{DO}{D. Oprisa and S. Stieberger, ``Six Gluon Open Superstring
Disk Amplitude, Multiple Hypergeometric Series and Euler-Zagier
Sums,'' arXiv: hep-th/0509042.}

















\end{thebibliography}
\end{document}